\documentclass[aps,prb,twocolumn,amsmath,amssymb,superscriptaddress,floatfix]{revtex4-2}

\usepackage[T1]{fontenc}
\usepackage[english]{babel}
\usepackage{graphicx}
\usepackage{booktabs}
\usepackage{bm}
\usepackage{physics}
\usepackage{hyperref}
\usepackage{xcolor}
\usepackage{placeins} % Prevents figures from floating into references

\graphicspath{{imagenes_en/}}

\begin{document}

\title{Band structure of two-dimensional graphene and of one-dimensional nanoribbons via
variational quantum algorithms: from energies to wavefunction quality}

\author{Ra\'ul Medrano Mill\'an}
\affiliation{Universidad Internacional de La Rioja (UNIR), Avenida de la Paz 137, 26006 Logro\~no, Spain}
\author{Rodrigo Gil-Merino y Rubio}
\affiliation{Universidad Internacional de La Rioja (UNIR), Avenida de la Paz 137, 26006 Logro\~no, Spain}
\author{Juan Borge}
\affiliation{Fisika Aplikatua Saila, Gipuzkoako Ingeniaritza Eskola, University of the Basque Country (EHU), Europa Plaza 1, 20018 Donostia/San Sebasti\'an, Spain}
\affiliation{Centro de F\'isica de Materiales (CFM-MPC), CSIC-UPV/EHU, 20018 Donostia/San Sebasti\'an, Spain}
\author{Unai Aseguinolaza Aguirreche}
\affiliation{Basic Sciences Department, Faculty of Engineering, Mondragon Unibertsitatea, 20500 Arrasate, Spain}

\date{\today}

\begin{abstract}
Simulating quantum many-body systems on classical computers encounters exponential computational bottlenecks, which quantum devices circumvent by natively encoding physical wavefunctions. In the Noisy Intermediate-Scale Quantum (NISQ) era, however, hardware noise severely constrains simulation accuracy. As a stepping stone toward simulating strongly correlated materials, we compute the band structure of two non-interacting graphene systems across different dimensions, the two-dimensional (2D) monolayer and one-dimensional (1D) nanoribbons with zigzag and armchair edges, using a single variational workflow in the $p_z$ tight-binding approximation. Beyond establishing a baseline, non-interacting ground states provide essential initial reference states for subspace and warm-started variational methods to solve interacting many-body spectra. By mapping the Bloch Hamiltonian to Pauli operators, we extract the ground-state band via the Variational Quantum Eigensolver and excited subbands via Variational Quantum Deflation, validating our approach from ideal simulations to execution on physical IBM Quantum processors. In ideal simulations, the workflow reproduces exact diagonalization results with high precision, whereas hardware noise introduces energy deviations of a few hundredths of the hopping parameter. While qubit requirements scale favorably with nanoribbon width, fixed-depth ansatz accuracy degrades with Hilbert space dimension. Crucially, deepening the ansatz improves ideal simulation accuracy by over three orders of magnitude but worsens energy estimates under noise, proving that practical circuit depth is bounded by hardware noise rather than physical expressivity. Finally, we complement energy evaluations with a wavefunction-level diagnostic measuring subspace containment. This diagnostic captures degenerate manifolds and exposes optimization failure modes that energy agreement alone disguises. Operating directly on NISQ hardware, this validated workflow establishes a controlled baseline for studying topological and strongly correlated phases in realistic 1D and 2D materials.
\end{abstract}

\maketitle

\section{Introduction}

Strongly correlated electron materials represent a critical paradigm for future technologies due to the emergent physical phenomena driven by dominant electron-electron interactions\cite{morosan2012strongly}. Unlike conventional metals and semiconductors described by non-interacting band theory, strong Coulomb repulsion yields collective phase transitions, including metal-insulator transitions\cite{mott1949basis,hubbard1963electron,imada1998metal}, unconventional superconductivity\cite{bednorz1986possible,scalapino2012common}, and spin-polarized ordering\cite{morosan2012strongly}. In carbon based architectures, electron-electron interactions drastically renormalize electronic structures driving superconductivity in magic-angle and crystalline multilayer graphene systems\cite{cao2018unconventional,pantaleon2023superconductivity}. In reduced dimensions, such as graphene nanoribbons (GNRs), enhanced Coulomb repulsion and quantum confinement induce localized edge states along zigzag boundaries\cite{son2006energy}. Furthermore, periodic structural engineering of GNRs gives rise to topological edge states and localized end states governed by symmetry-protected topological phases\cite{cao2017topological,groning2018engineering}. Harnessing these interaction-driven superconducting phases alongside localized and topological boundary states require many-body accounting of electronic band structures to guide material engineering at the atomic scale.

Simulating strongly correlated many-body quantum physics on classical computers encounters severe computational bottlenecks. Exact diagonalization explicitly solves the $N$-body Schödinger equation, but is strictly constrained to tiny clusters due to the exponential growth of Hilbert space dimension. Tensor network methods like Density Matrix Renormalization Group (DMRG)\cite{schollwock2011density}, deliver near-exact precision for quasi-one-dimensional systems by truncating low-entanglement states, however, extending DMRG to higher dimensional lattices requires an exponential expansion in bond dimension. Finally, Green's function approximations and mean-field extensions such as $GW$\cite{aryasetiawan1998gw} and Dynamical Mean-Field Theory\cite{georges1996dynamical} simplify electron correlations by mapping lattice interactions onto local impurity model, yet, they fail to capture non-local correlation fluctuations, long-range entanglement, and topological edge states without empirical parametrization. 

Quantum computers circumvent classical simulation bottlenecks by using controllable quantum states to store and process physical wavefunctions\cite{Feynman1982}. By mapping fermionic operators onto qubit registers\cite{bravyi2002fermionic,jordan1928paulische}, a system of $N$ spin-orbitals maps directly onto $N$ qubits, storing the full $2^{N}$-dimensional Hilbert space. Algorithms such as Quantum Phase Estimation (QPE)\cite{nielsen2000quantum} leverage unitary time evolution to calculate ground-state energies and energy bands with polynomial circuit scaling. However, executing long-depth algorithms such as QPE remains infeasible on current devices. In the present NISQ\cite{Preskill2018} era, hardware is constrained by gate errors, crosstalk, and limited coherence times. To overcome these limitations hybrid quantum-classical approaches like the Variational Quantum Eigensolver (VQE)\cite{McClean2016,Peruzzo2014} have emerged. By parametrizing low-depth quantum circuits and performing parameter optimization in classical hardware, variational algorithms minimize coherence requirements providing a practical route to calculate electron correlations and band structures\cite{Bharti2022,Bauer2020,ma2020quantum,bassman2021simulating}. Furthermore, quantum strategies and scalable algorithms for correlated lattice systems such as the Fermi-Hubbard model\cite{cade2020strategies,stanisic2022observing} and demonstrations of quantum utility\cite{kim2023evidence} have expanded the reach of near-term processors in condensed matter physics.

Although the ultimate goal of quantum computing in materials science is to simulate strongly correlated many-body Hamiltonians beyond classical reach, implementing non-interacting tight-binding Hamiltonians on physical processors represents more than a preliminary step. Subspace algorithms, such as Quantum Krylov Subspace (QKS) methods, Krylov Quantum Diagonalization (KQD) \cite{cortes2022quantum}, and Sample-Based Krylov Quantum Diagonalization (SKQD) \cite{yu2025quantum}, critically depend on high-quality initial reference states, as their convergence rates and subspace sampling efficiencies rely on the overlap with the true low-energy manifold. Likewise, warm-started Variational Quantum Eigensolver\cite{egger2021warm,puig2025variational} protocols leverage single-particle approximations to initialize variational parameters in favorable regions of the Hilbert space, mitigating barren plateaus and accelerating convergence toward correlated ground states. Single-particle tight-binding solutions offer a controlled starting point to build time-evolved Krylov bases or sample relevant electronic configurations or warm-start variational optimizers. In this context, Cerasoli et al. \cite{cerasoli2020quantum} pioneered mapping empirical tight-binding Hamiltonians of bulk silicon onto Pauli operators to solve for energy bands on IBM Quantum hardware, while Sherbert et al. and others \cite{Sherbert2022band,zhang2024full} established a generalized framework for calculating band structures from atomic orbitals. However, applying these algorithms to low-dimensional architectures such as one-dimensional graphene nanoribbons (GNRs) on physical quantum hardware remains largely unexplored.

In this work we address two carbon systems of different dimensionality within a single variational framework. On the one hand we take the two-dimensional monolayer graphene. On the other hand we take the quasi-one-dimensional GNRs. First, we perform the ideal simulation of the proposed method and we recover the exact diagonalization results. Second we perform simulations under quantum hardware noise models and we see that finite sampling and noisy optimization have comparable contributions. Third we perform executions in real IBM Quantum processors and we demonstrate the feasibility of the computation on current NISQ devices. We also add a subspace containment analysis to adress not only the energy, but also the wavefunction quality. 

The work is organized as follows: In section \ref{sec:modelo} we define the tigh binding model of graphene and GNRs. In section \ref{sec:metodos} we define the quantum algorithm we will use to obtain the band structures of graphene and GNRs. In sections \ref{sec:res_monocapa} and \ref{sec:res_nanocintas} we show the band structure calculations. In section \ref{sec:containment} we show the results regarding subspace containment. Finally, we explain the conclusions in section \ref{sec:conclusions}.

% ------------------------------------------------------------------

\section{Theoretical model}
\label{sec:modelo}

This section introduces the two systems studied in this work and the Hamiltonians from which their bands are obtained. We first describe two-dimensional monolayer graphene and then the quasi-one-dimensional nanoribbons, in both cases within the same tight-binding framework and in the Bloch representation, which reduces the problem, at each point of reciprocal space, to a finite-dimensional matrix $H(\bm{k})$. Throughout the paper, the exact reference against which every variational result is compared is the direct numerical diagonalization (exact diagonalization) of that Bloch matrix on a dense sampling of $\bm{k}$; in all band figures (Figs.~\ref{fig:bandas_monocapa}--\ref{fig:hardware}), the exact diagonalization result is drawn as continuous gray lines and the variational result as colored points.

\begin{figure*}[t]
  \centering
  \includegraphics[width=\textwidth]{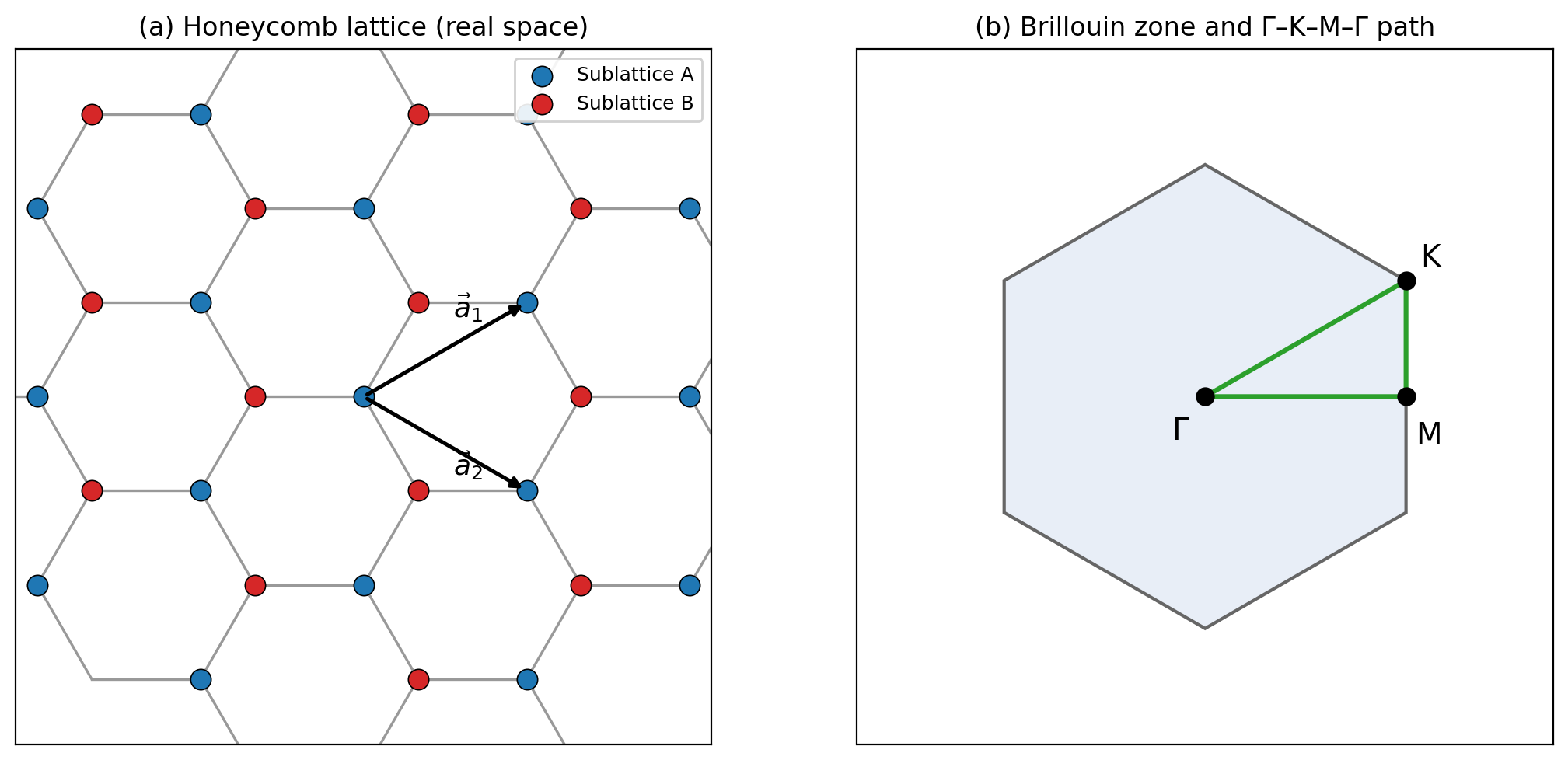}
  \caption{Structure of monolayer graphene. (a)~Honeycomb lattice in real space, with the sublattices $A$ and $B$ and the lattice vectors $\vec{a}_1=\frac{a}{2}\left(3,\sqrt{3}\right),\vec{a}_2=\frac{a}{2}\left(3,-\sqrt{3}\right)$. (b)~First Brillouin zone (hexagon) with the high-symmetry points $\Gamma$, $K$ and $M$ and the path $\Gamma\!\to\!K\!\to\!M\!\to\!\Gamma$. Own elaboration.}
  \label{fig:monocapa}
\end{figure*}

Graphene is a hexagonal (honeycomb) lattice of carbon atoms with two triangular sublattices $A$ and $B$ (Fig.~\ref{fig:monocapa}a). In the tight-binding approximation, the $\pi$ electrons are described by
\begin{equation}
  H = -t \sum_{\langle i,j\rangle} c_i^\dagger c_j - t' \sum_{\langle\langle i,j\rangle\rangle}
  c_i^\dagger c_j + \mathrm{h.c.},
\end{equation}
where $t$ is the first-nearest-neighbor hopping integral and $t'$ the second-nearest-neighbor one ($t'\!\approx t/28$). By periodicity, for each wave vector $\bm{k}$ the problem reduces to a $2\times 2$ Bloch Hamiltonian over the two sublattices,
\begin{equation}
  H(\bm{k}) = \begin{pmatrix} c(\bm{k}) & -t\,f(\bm{k}) \\ -t\,f^{\ast}(\bm{k}) & c(\bm{k}) \end{pmatrix},
  \label{eq:H2x2}
\end{equation}
with
\begin{equation}
  f(\bm{k}) = \sum_{\bm\delta} e^{\,i\bm{k}\cdot\bm\delta},
  \qquad
  c(\bm{k}) = 2t' \sum_{\bm\delta'} \cos(\bm{k}\cdot\bm\delta'),
\end{equation}
where $\{\bm\delta\}$ are the three first-nearest-neighbor vectors and $\{\bm\delta'\}$ the six second-nearest-neighbor ones. Their energies are $E_{\pm}(\bm{k}) = c(\bm{k}) \pm t\,|f(\bm{k})|$. At first nearest neighbors ($t'=0$) the diagonal term vanishes, $c=0$, and the bands are symmetric about the Fermi level, $E_{\pm}=\pm t|f|$, touching at the Dirac points $\bm{K}$ where $f(\bm{K})=0$; the second-nearest-neighbor term $c(\bm{k})$ breaks that electron-hole symmetry and shifts the Dirac point in energy. The band structure is traced along the high-symmetry path $\Gamma\!\to\! K\!\to\! M\!\to\!\Gamma$ of the first Brillouin zone (Fig.~\ref{fig:monocapa}b).

\begin{figure}[t]
  \centering
  \includegraphics[width=\columnwidth]{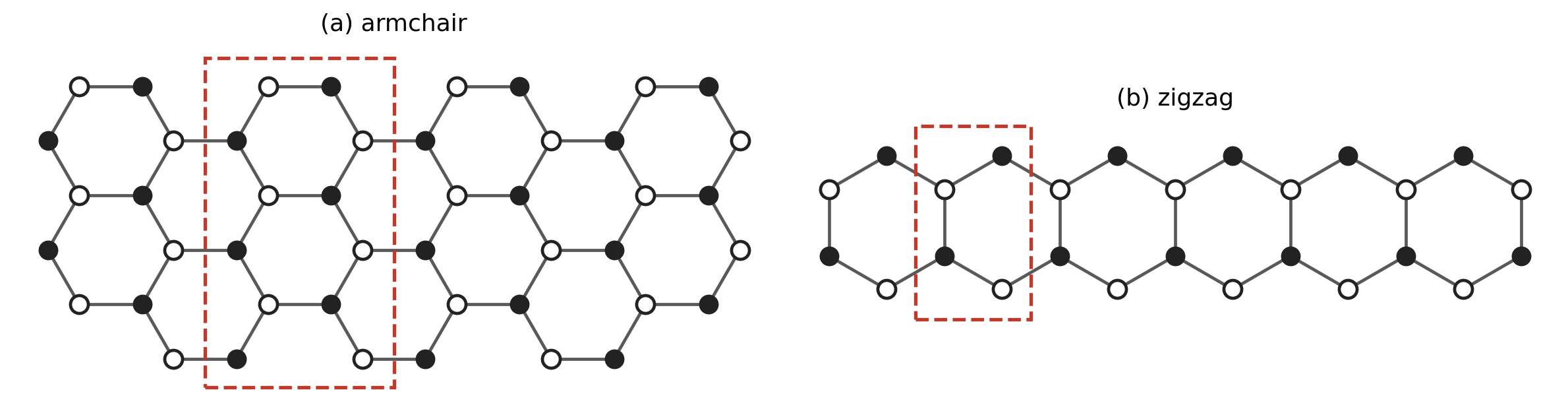}
  \caption{Structure of graphene nanoribbons with armchair (a) and zigzag (b) edges. Filled and open circles represent the two sublattices ($A$ and $B$), and the dashed rectangle indicates the unit cell. Own elaboration.}
  \label{fig:cintas}
\end{figure}

We turn now to the one-dimensional systems. Confining the sheet yields a system that is periodic along a single direction (the ribbon axis), with a unit cell of $N$ atoms per sublattice. The Bloch Hamiltonian preserves the block structure
\begin{equation}
  H(k) = \begin{pmatrix} 0 & f(k) \\ f^{\dagger}(k) & 0 \end{pmatrix},
  \label{eq:Hblock}
\end{equation}
where now $f(k)$ is an $N\times N$ matrix and $k$ is the wave vector along the ribbon. The form of $f(k)$ depends on the type of edge~\cite{Wakabayashi2010} (Fig. \ref{fig:cintas}):
\begin{itemize}
  \item \textbf{Zigzag:} main diagonal $-t\,(1 + e^{ik})$; subdiagonal $-t$.
  \item \textbf{Armchair:} main diagonal $-t\,e^{ik/2}$; subdiagonal and superdiagonal $-t$.
\end{itemize}
The full Hamiltonian $H(k)$ has dimension $2N\times 2N$; its $2N$ eigenvalues at each $k$ constitute the bands of the ribbon. In the implementation (Sec.~\ref{sec:metodos}) we take $N = 2^{n_q-1}$, so that $H(k)$ has dimension $2^{n_q}\times 2^{n_q}$ and is represented with exactly $n_q$ qubits.

\section{Quantum methods}
\label{sec:metodos}

This section describes the computational workflow, common to both systems, that takes each Bloch Hamiltonian to a band structure obtained on a quantum processor. We first explain how the matrix $H(\bm{k})$ is rewritten as an operator acting on qubits and which parametrized circuit prepares the trial state; we then describe the VQE and VQD algorithms used to obtain, respectively, the lowest band and the excited bands. We next define the three execution modes in which every calculation of this work can be run, discuss the computational cost of the noise simulation and the error channels it contains, and present the strategy of execution on real hardware. Finally, we introduce the subspace-containment diagnostic with which we validate the variational wavefunction.

For each $\bm{k}$ the matrix $H(\bm{k})$ of dimension $2^{n_q}\times 2^{n_q}$ is constructed and decomposed in the basis of Pauli-matrix products,
\begin{equation}
  H(\bm{k}) = \sum_{\alpha} c_\alpha(\bm{k})\, P_\alpha, \qquad P_\alpha \in
  \{I,X,Y,Z\}^{\otimes n_q},
\end{equation}
an operator acting on $n_q$ qubits. The trial state on which this operator is evaluated is prepared with a parametrized circuit: we use the hardware-efficient ansatz EfficientSU2 (with a number of repetitions that is stated for each calculation), on account of its low gate count and its suitability for NISQ devices.

The lowest-energy band at each $\bm{k}$ is obtained with VQE, by minimizing the expectation value
\begin{equation}
  E_0(\bm{k}) = \min_{\bm\theta}\ \langle \psi(\bm\theta)|\,H(\bm{k})\,|\psi(\bm\theta)\rangle,
\end{equation}
by means of a classical optimizer. We have employed COBYLA, SPSA and SLSQP: SPSA is more robust against the stochastic noise of the measurement~\cite{Spall1992,Kandala2017} (preferable with finite shots and on real hardware), whereas SLSQP, being a gradient-based method, converges rapidly and accurately in ideal simulations.

The upper bands are obtained with VQD, by deflation: when computing the $j$-th state, a penalty term is added to the cost function that enforces orthogonality with respect to the states already found,
\begin{equation}
  C_j(\bm\theta) = \langle \psi(\bm\theta)|H(\bm{k})|\psi(\bm\theta)\rangle
  + \beta \sum_{l<j} \big| \langle \psi(\bm\theta)|\psi_l\rangle \big|^2 ,
  \label{eq:vqd}
\end{equation}
with a hyperparameter $\beta$, set to $\beta=10$ throughout this work, large compared with the bandwidth $\sim 6|t|$, so that the penalty dominates over any energetic gain from collapsing onto an already occupied level. To accelerate convergence along the path in $\bm{k}$ we use the optimal parameters of one point $\bm{k}$ as the seed for the next (warm-start). Two practical remarks about this deflation loop are worth recording. The first is that the sequential character of VQD makes the $j$-th state inherit the errors of the $j-1$ preceding ones and accumulate $j$ penalty terms, so that the quality of the reconstruction degrades systematically with the index of the state. The second is that, on a noisy cost function, a local optimizer started from a single seed can settle in a local minimum in which a state skips a level, it converges onto an eigenspace that is not its own, while the deflation penalty keeps it orthogonal to the previous states. We found that increasing the iteration budget does not correct this and can even aggravate it, since on a noisy cost function the optimizer keeps wandering around the wrong minimum, whereas restarting the optimization from a different seed does recover the correct eigenspace. We therefore optimize each state from $n_r$ seeds, the first being the warm-start one, the rest random, and retain the one of lowest cost, evaluated on the ideal simulator so that the choice does not depend on sampling noise. Unless stated otherwise, we use $n_r=1$; $n_r=3$ is used in the three-qubit containment calculations of Sec.~\ref{sec:containment}, where the fidelity of the excited states is itself the quantity under study.

Three execution modes are used throughout the paper:
\begin{itemize}
  \item statevector: ideal simulation. The expectation value is evaluated exactly on the state vector, without sampling or noise.
  \item aer\_noise: simulation with a noise model of a device (see below), with a finite number of shots.
  \item hardware: execution on a real IBM Quantum processor.
\end{itemize}
We do not add the ideal simulation with a finite number of shots. The reason is that the statistical error of a finite sampling is already contained in aer\_noise, which adds the device errors on top of it, and, in the systems studied here, the results are indistinguishable from statevector within that statistical error. The decomposition of the error into its sampling and optimization contributions, reported below, is obtained instead from the two energies recorded in every mode: the ideal one (exact expectation value over the optimal parameters) and the measured one (with noise or hardware), which allows separating the robustness of the optimizer from the effect of noise on the energy.

For the aer\_noise simulations we adopt the systematic rule of emulating a device with exactly the number of qubits required by the circuit with a fixed seed for reproducibility. This procedure can reduce the computational cost by a factor of 500 compared to the simulation using a simulated backend of 133 qubits and does not alter the noise model.

The emulated device has the native gate set $\{R_z, \sqrt{X}, X, \mathrm{CX}\}$, and the noise model assigns to each of them a depolarizing channel with the corresponding gate error, composed with a thermal relaxation channel built from the gate duration and the coherence times $T_1$ and $T_2$ of the qubits involved; the measurement carries, in addition, a classical readout error (bit-flip on the recorded outcome). The virtual $R_z$ rotations are exact and carry no error, as on the real device. For the sizes used here the model yields one-qubit gate errors of $\sim\!10^{-4}$ (durations of $30$--$60$~ns), two-qubit CX errors of $\sim\!1$--$4\times10^{-3}$ (durations of $0.3$--$0.9$~$\mu$s), readout errors of $\sim\!3$--$5\times10^{-3}$, and coherence times $T_1, T_2 \approx 120$--$200$~$\mu$s, values representative of the current generation of superconducting devices. Not included are correlated errors, crosstalk between neighboring qubits and the temporal drift of the calibration, all of which are present on the real machine. The model is therefore expected to underestimate the error with which a given circuit is measured on the device, mainly through the errors associated with two-qubit gates; Sec.~\ref{sec:res_nanocintas} quantifies this for the two-qubit circuits of the nanoribbons, while the single-qubit circuit of the monolayer, which contains no two-qubit gates, is essentially unaffected (Sec.~\ref{sec:res_monocapa}).

Running the variational optimization directly on hardware consumes a lot of quantum computing time because the cost function must be evaluated at each step. We therefore adopt, in the hardware mode, a hybrid strategy: the ansatz parameters are optimized entirely in simulation and only the optimal circuit of each band and $\bm{k}$ point is run on the real device to measure its energy. In this way the number of jobs is reduced to one per (band, $\bm{k}$) pair: the hardware-measured energy of the optimal variational state. In all hardware calculations readout error mitigation is applied. This strategy, which we refer to below as the hybrid hardware strategy, is the one used in every real-device result reported in this work.

The agreement in energy does not by itself guaranty that the state prepared by the circuit is the correct physical state. A variational state can approach the exact eigenvalue while carrying appreciable weight in neighboring eigenspaces. To validate the wavefunction we adopt the subspace containment diagnostic introduced in Ref.~\cite{Exposito2026FQHE}. For a variational state $|\psi(\bm\theta)\rangle$ and an exact-diagonalization eigenspace $\mathcal{H}_\alpha$, we define the subspace containment
\begin{equation}
  C_\alpha(\bm\theta) = \langle \psi(\bm\theta)|\,\hat{P}_\alpha\,|\psi(\bm\theta)\rangle ,
  \label{eq:containment}
\end{equation}
where $\hat{P}_\alpha$ is the projector onto $\mathcal{H}_\alpha$. If $\{|\phi^{\mathrm{ED}}_{\alpha,i}\rangle\}$ is an orthonormal basis of that eigenspace, this becomes
\begin{equation}
  C_\alpha(\bm\theta) = \sum_{i} \big| \langle
  \phi^{\mathrm{ED}}_{\alpha,i}|\psi(\bm\theta)\rangle \big|^2 .
  \label{eq:containment2}
\end{equation}

Eq.~\eqref{eq:containment2} covers two cases. For a nondegenerate eigenstate the eigenspace is one-dimensional and the containment reduces to the usual squared overlap with the corresponding exact eigenvector: $C_\alpha = |\langle\phi^{\mathrm{ED}}_\alpha|\psi(\bm\theta)\rangle|^2$, a number that says what fraction of the variational state is the physical state one intended to prepare, with $1-C_\alpha$ the weight leaked into the remaining eigenvectors. For a degenerate manifold, by contrast, the projection onto the full eigenspace is the appropriate quantity because it is basis-independent. In that case the $d$ variational states obtained sequentially by VQD within a $d$-fold degenerate manifold should be understood as $d$ states spanning the same eigenspace, not as energetically distinct levels, and what is meaningful is that each of them and, jointly, the subspace they span, resides within $\mathcal{H}_\alpha$.

In the systems studied here degeneracies are confined to isolated high-symmetry points: the Dirac point $K$ of the monolayer, where the two $\pi$ bands touch. The zone boundary $k=\pi$ of the zigzag ribbon, where the spectrum collapses onto three eigenspaces: $E=-|t|$ and $E=+|t|$, each $(N-1)$-fold degenerate, so that their degeneracy grows with the width, and the twofold eigenspace of the two edge states at $E=0$. Finally, the degeneracies of the armchair ribbon at $k=2\pi/3$ and $k=\pi$ (see Fig. \ref{fig:anchura}). At those points, the containment must be read over the manifold in the sense just described. Eigenvalues are grouped into eigenspaces numerically, with a tolerance $\mathrm{tol}_{\mathrm{deg}}=10^{-6}$ in units of $|t|$, so that the two regimes are separated automatically at each $\bm{k}$.

The containment is evaluated on the noiseless statevector of the optimal parameters and on parameters obtained in the presence of noise. 

\section{Results: monolayer graphene (2D)}
\label{sec:res_monocapa}

This section reports the band structure of monolayer graphene along the path $\Gamma\!\to\!K\!\to\!M\!\to\!\Gamma$. Being a single-qubit system, the monolayer is the simplest case and sets the reference against which the nanoribbons of Sec.~\ref{sec:res_nanocintas} are read.

\begin{figure*}[t]
  \centering
  \includegraphics[width=\textwidth]{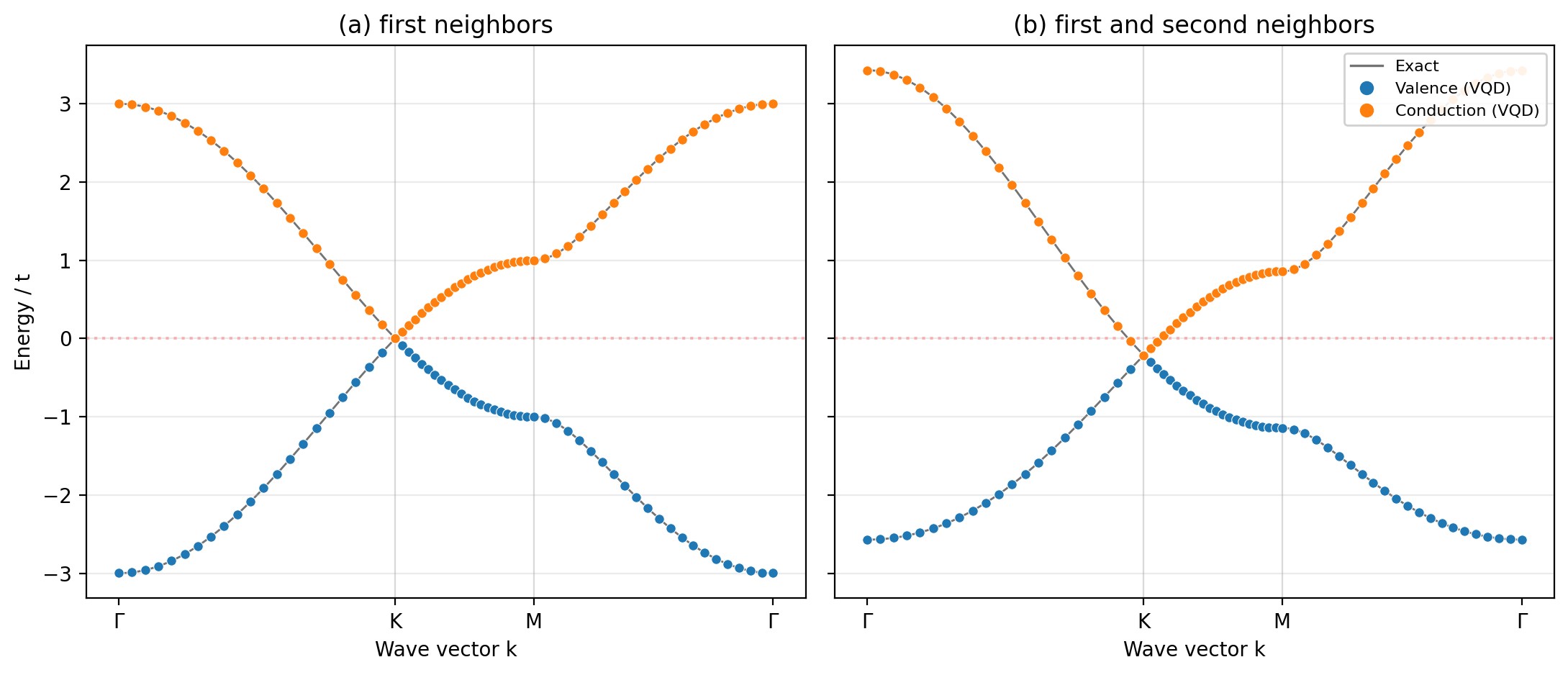}
  \caption{Band structure of monolayer graphene in the statevector mode, obtained by VQD (valence in blue, conduction in orange) versus exact diagonalization (gray lines), along the path $\Gamma\!\to\!K\!\to\!M\!\to\!\Gamma$: (a)~with first nearest neighbors and (b)~with first and second nearest neighbors. SLSQP optimizer, EfficientSU2 ansatz with two repetitions.}
  \label{fig:bandas_monocapa}
\end{figure*}

Fig.~\ref{fig:bandas_monocapa} compares the valence and conduction bands obtained with exact diagonalization and statevector mode, in the two approximations of the model. At first nearest neighbors (Fig.~\ref{fig:bandas_monocapa}a) the bands are symmetric about the Fermi level and touch at the Dirac point $K$; upon including the second nearest neighbors (Fig.~\ref{fig:bandas_monocapa}b) the electron-hole symmetry is broken and the Dirac point shifts in energy. In both cases the agreement with the exact reference is excellent along the entire path, which validates the VQE/VQD workflow for the two-dimensional system.

\begin{figure*}[t]
  \centering
  \includegraphics[width=\textwidth]{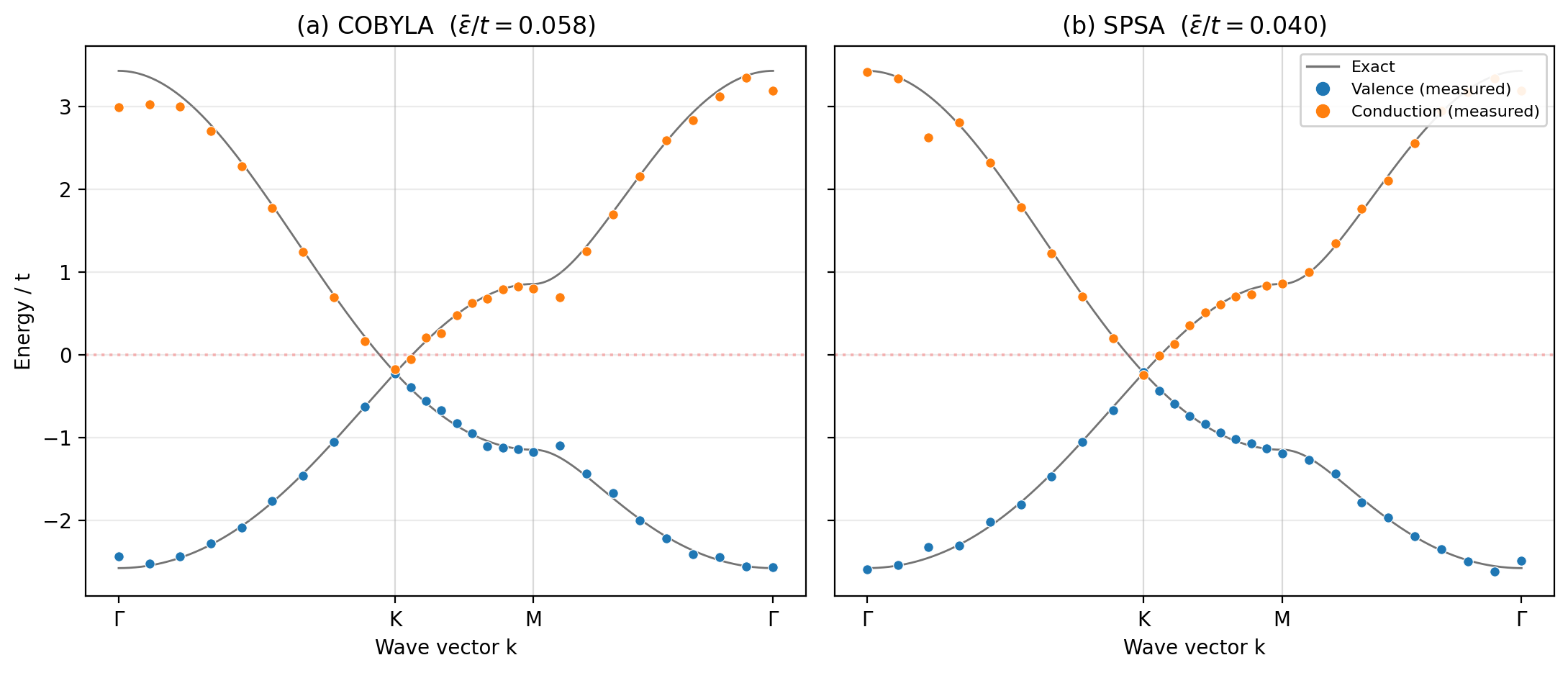}
  \caption{Effect of noise in monolayer graphene (aer\_noise mode): energy measured under noise (valence in blue, conduction in orange) versus exact diagonalization (gray lines), along $\Gamma\!\to\!K\!\to\!M\!\to\!\Gamma$, optimizing with (a)~COBYLA and (b)~SPSA. The mean error $\bar\epsilon/t$ with respect to the exact result is indicated.}
  \label{fig:ruido_monocapa}
\end{figure*}

Fig.~\ref{fig:ruido_monocapa} compares the energy measured under noise with exact diagonalization. The optimizations are done with COBYLA and SPSA. In both cases the band structure is well reproduced, with mean deviations of $\bar\epsilon/t \approx 5.8\times 10^{-2}$ (COBYLA) and $\bar\epsilon/t \approx 4.0\times 10^{-2}$ (SPSA). The SPSA optimizer turns out to be more robust on average against the stochastic noise of the measurement, although both exhibit larger pointwise deviations in some regions of the path, in line with systematic comparisons showing that the relative performance of these optimizers is strongly problem dependent~\cite{BonetMonroig2023}. These deviations are consistent with the measurement noise level associated with a finite sampling of 1024 shots.

\begin{figure}[t]
  \centering
  \includegraphics[width=\columnwidth]{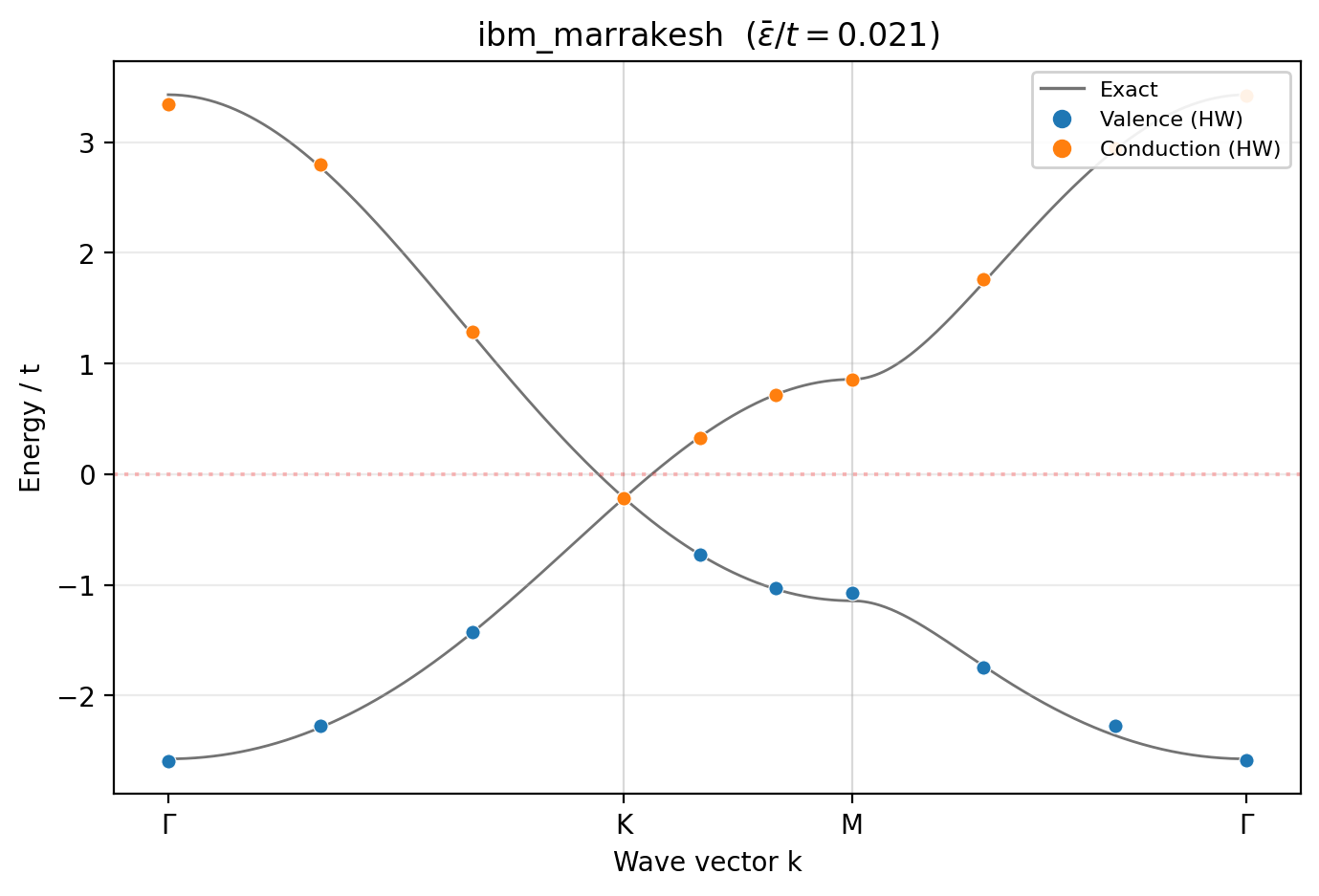}
  \caption{Band structure of monolayer graphene computed on a real IBM Quantum computer (ibm\_marrakesh, Heron family): energy measured on the hardware (points: valence in blue, conduction in orange) versus exact diagonalization (gray lines), along $\Gamma\!\to\!K\!\to\!M\!\to\!\Gamma$. Execution in the hardware mode with the hybrid strategy of Sec.~\ref{sec:metodos}. The mean error $\bar\epsilon/t$ is indicated.}
  \label{fig:hw_monocapa}
\end{figure}

Fig.~\ref{fig:hw_monocapa} compares the energy measured on the hardware with exact diagonalization along the same path. The two bands are reproduced with a mean deviation of $\bar\epsilon/t \approx 2.1\times 10^{-2}$ including the Dirac point at $K$, an excellent result that reflects that the circuit is shallow and the impact of device noise is small.

This error is smaller than the one obtained with the noise model in Fig.~\ref{fig:ruido_monocapa}, which might seem to contradict the expectation that the model underestimates device noise (Sec.~\ref{sec:metodos}). However, the two values measure different quantities. In aer\_noise mode, parameters are optimized under noise, meaning the total error $\bar\epsilon$ includes two distinct components: optimization error ($2.2\times10^{-2}$ for SPSA; $4.5\times10^{-2}$ for COBYLA) and final measurement error ($2.4\times10^{-2}$ for SPSA; $2.5\times10^{-2}$ for COBYLA). In contrast, the hybrid hardware strategy optimizes parameters in an ideal simulation, reaching exact energies within $10^{-7}\,t$, so its reported error consists solely of the final measurement component. When isolating pure measurement error, the physical device ($2.1\times10^{-2}$) and the noise model ($2.4\text{--}2.5\times10^{-2}$) agree closely. The device does not underperform the model here because this single-qubit circuit involves no two-qubit gates, eliminating unmodeled hardware noise like crosstalk and correlated errors, while single-qubit gate errors are negligible ($\sim 10^{-4}$). Furthermore, readout error mitigation was applied on the real hardware but omitted in simulation. As a result, the remaining error in both cases is dominated simply by finite sampling noise.

\section{Results: graphene nanoribbons (1D)}
\label{sec:res_nanocintas}

We turn now to the one-dimensional systems. This section reports, in order: (i)~the validation of the method in statevector mode; (ii)~the dependence of the spectrum on the ribbon width, obtained by exact diagonalization; (iii)~the effect of the device noise model, including the decomposition of the error into its sampling and optimization contributions; (iv)~the execution on real IBM Quantum processors; and (v)~a scalability and error analysis as the number of qubits is increased, carried out in statevector and aer\_noise modes.

\begin{figure*}[t]
  \centering
  \includegraphics[width=\textwidth]{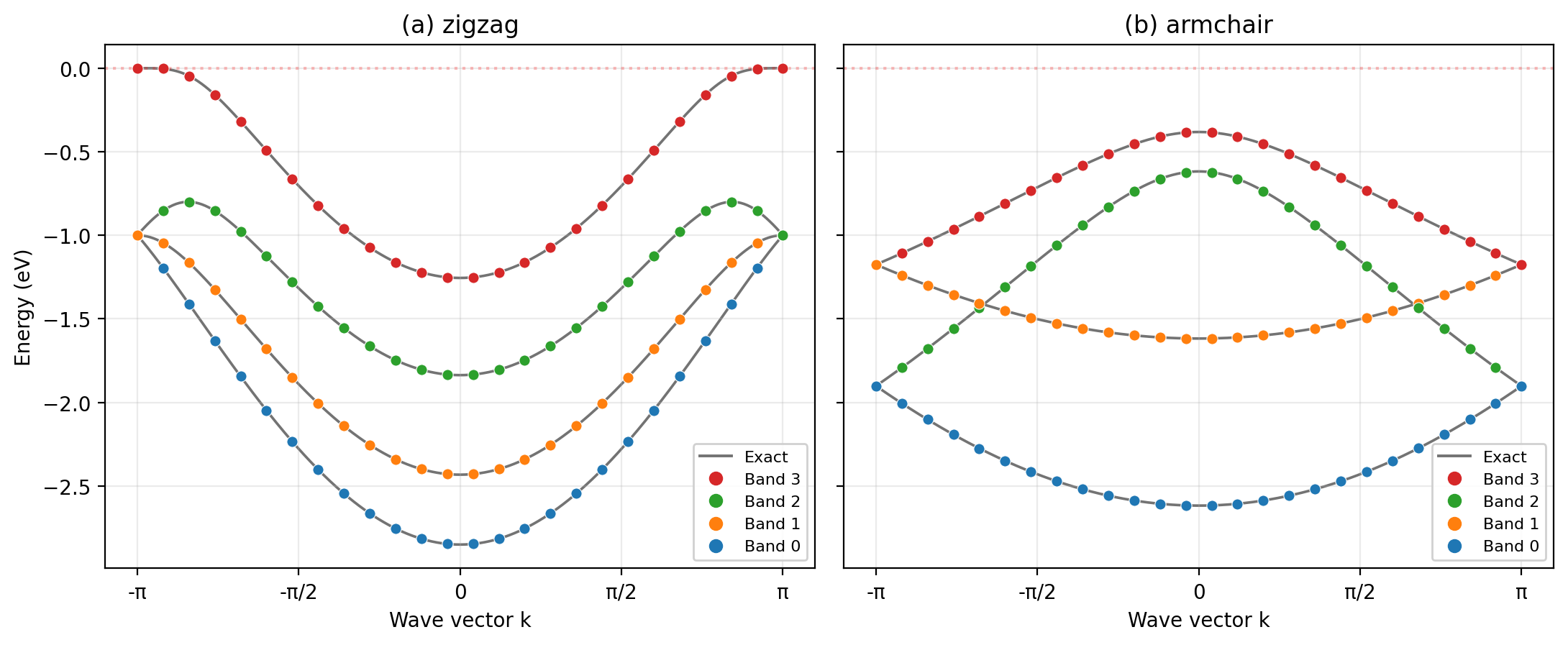}
  \caption{Band structure of GNBs ($n_q=3$, $N=4$) obtained in statevector mode by VQD (colored points) versus exact diagonalization (gray lines), for the four lowest bands, in the (a) zigzag and (b) armchair ribbons, over $k\in[-\pi,\pi]$. SLSQP optimizer and EfficientSU2 ansatz with two repetitions.}
  \label{fig:validacion}
\end{figure*}

\begin{figure*}[t]
  \centering
  \includegraphics[width=\textwidth]{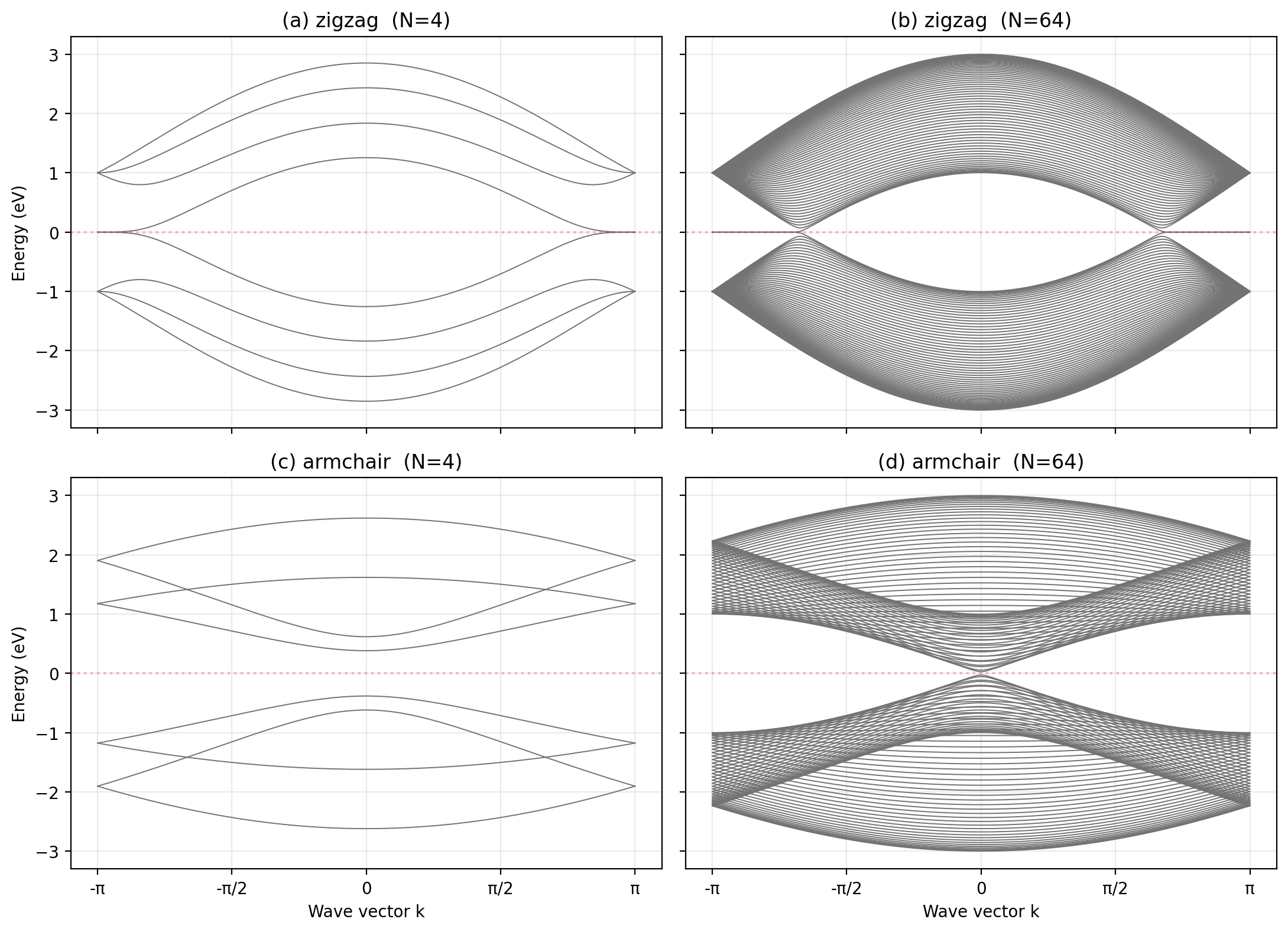}
  \caption{Exact band structure as a function of the ribbon width, obtained by exact diagonalization: zigzag with (a) $N=4$ and (b) $N=64$, and armchair with (c) $N=4$ and (d) $N=64$.}
  \label{fig:anchura}
\end{figure*}

Fig.~\ref{fig:validacion} compares the statevector results of GNBs with exact diagonalization. The band structure of both ribbons with $n_q=3$ (width $N=4$) over the entire interval $k\in[-\pi,\pi]$ is optimized with SLSQP using an EfficientSU2 ansatz with two repetitions. The agreement is excellent for the four bands in both ribbons. It is worth noting that reaching this precision in the excited bands required a sufficiently expressive ansatz (two repetitions) and a gradient-based optimizer (SLSQP); with a single-repetition ansatz the excited bands of the zigzag ribbon exhibit appreciable deviations.

We next examine how the spectrum depends on the width of the ribbon Fig.~\ref{fig:anchura} shows the exact band structure as the width is varied between $N=4$ ($n_q=3$) and $N=64$ ($n_q=7$). As $N$ increases, the number of subbands grows and the spectrum tends toward that of two-dimensional graphene. In the zigzag ribbons one can appreciate the nearly flat pair of edge-state bands at $E\approx0$ for $k\in[2\pi/3,\pi]$, which becomes flatter as the width increases. In the armchair ribbons a gap appears around $E=0$ that varies with $N$. It should be noted that, by construction of the mapping ($N=2^{n_q-1}$), the accessible widths are restricted to powers of two; a systematic study of the different width families of the armchair ribbons, with their well-known dependence of the gap on $N$~\cite{Son2006}, would require a more flexible mapping.

\begin{figure*}[t]
  \centering
  \includegraphics[width=\textwidth]{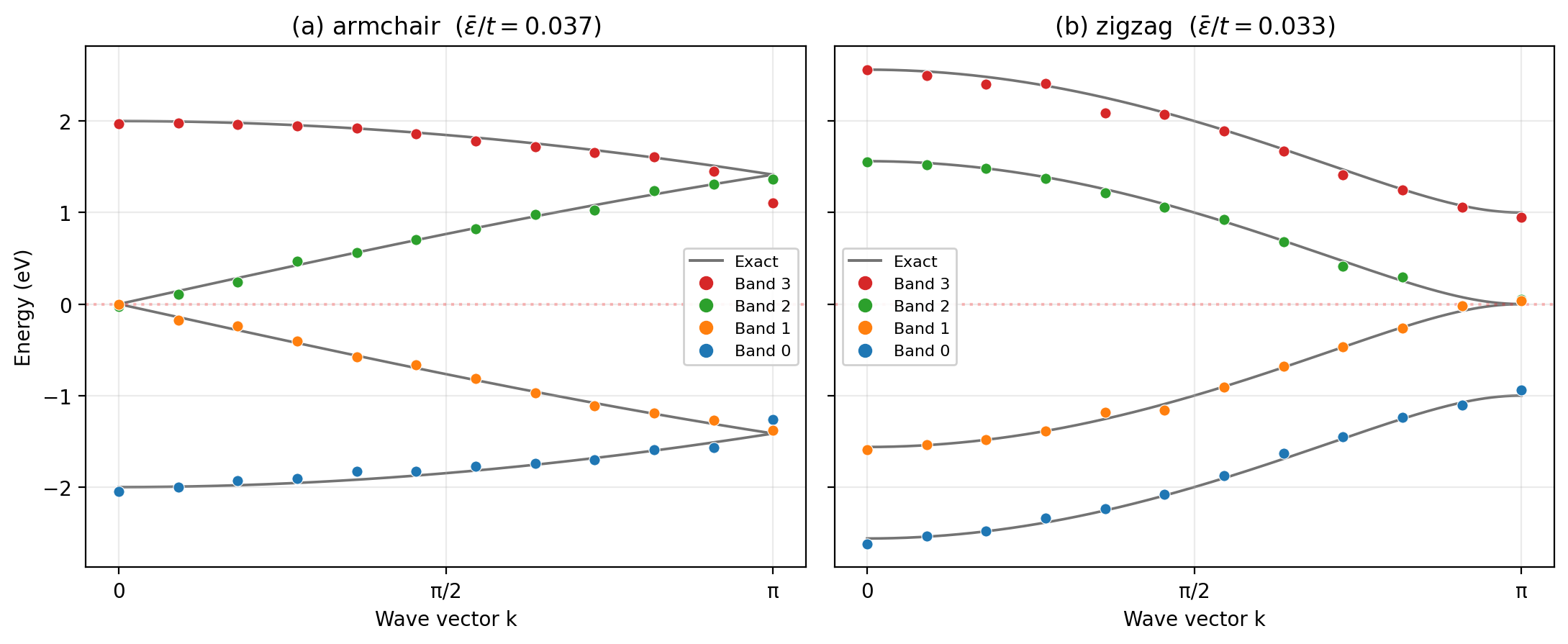}
  \caption{Simulation in the aer\_noise mode ($n_q=2$, $N=2$): energy measured under noise (colored points) versus exact diagonalization (gray lines), for the (a) armchair and (b) zigzag ribbons, over $k\in[0,\pi]$. Optimization with SPSA. The mean error $\bar\epsilon/t$ with respect to the exact result is indicated.}
  \label{fig:ruido}
\end{figure*}

To study the effect of noise we reproduce the band structure of both ribbons ($n_q=2$, four bands) in the aer\_noise mode with the SPSA optimizer. Fig.~\ref{fig:ruido} shows that the measured energy reproduces the exact band structure with good fidelity despite the noise. In units of $t$, the mean deviation with respect to the exact result is $\sim 3.7\times 10^{-2}$ in the armchair ribbon and $\sim 3.3\times 10^{-2}$ in the zigzag one (indicated in each panel of Fig.~\ref{fig:ruido}). The decomposition of the error into its two contributions reveals that both are comparable: the measurement noise (measured energy versus the ideal energy of the optimal parameters) is $\sim 2.5\times 10^{-2}$, consistent with the finite sampling of 1024 shots ($\sim 1/\sqrt{1024}$ per term), while the effect of noise on the optimization (ideal energy of the optimal parameters obtained with noise versus the exact one) is $\sim 1.5\times 10^{-2}$. Overall, the SPSA optimizer maintains good precision despite the noise.

\begin{figure*}[t]
  \centering
  \includegraphics[width=\textwidth]{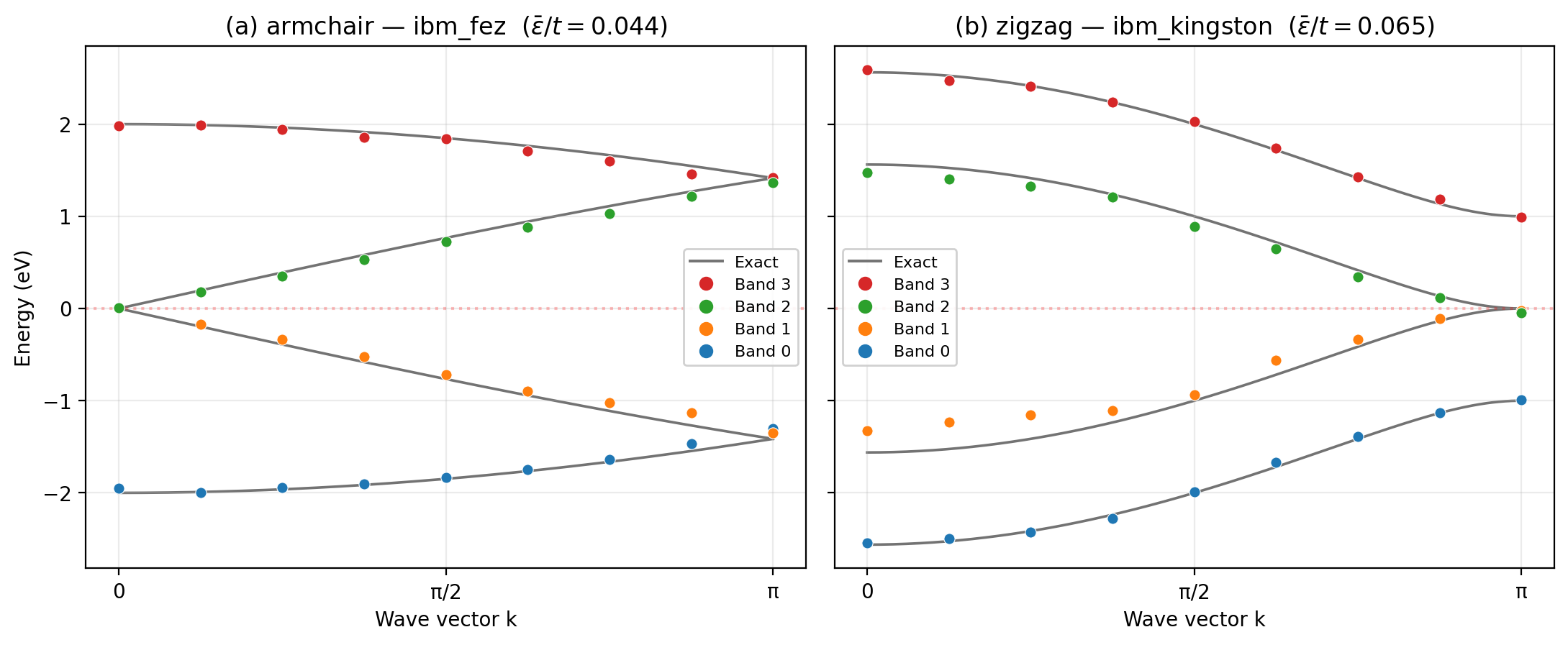}
  \caption{Band structure of graphene nanoribbons ($n_q=2$, $N=2$) computed on real IBM Quantum computers: energy measured on the hardware (colored points) versus exact diagonalization (gray lines), for the (a) armchair (backend ibm\_fez) and (b) zigzag (backend ibm\_kingston) ribbons, over $k\in[0,\pi]$. Execution in the hardware mode with the hybrid strategy of Sec.~\ref{sec:metodos}. The mean error $\bar\epsilon/t$ is indicated.}
  \label{fig:hardware}
\end{figure*}

Next, we compute the complete band structure of the narrowest nanoribbons ($n_q=2$, $N=2$) on real IBM Quantum computers. We used the processors ibm\_fez (armchair ribbon) and ibm\_kingston (zigzag ribbon), both of the IBM~Heron family. Fig.~\ref{fig:hardware} compares the energy measured on the hardware with exact diagonalization. In the armchair ribbon the four bands are reproduced with a mean deviation of $\bar\epsilon/t\approx4.4\times10^{-2}$, distributed fairly evenly among the bands ($3.7$--$5.9\times10^{-2}$). In the zigzag ribbon the mean deviation is $\bar\epsilon/t\approx6.5\times10^{-2}$: the lowest and highest bands are reproduced with $2.1\times10^{-2}$ and $2.3\times10^{-2}$, while the two central bands concentrate most of the error ($1.4\times10^{-1}$ and $7.2\times10^{-2}$), with deviations of up to $0.29\,t$. These larger deviations are not related to the near degeneracy of the central bands at $E\approx0$: they are largest at small $k$, where the central bands are well separated from the rest of the spectrum, and smallest close to $k=\pi$, where they become degenerate.

Because the parameters are optimized in ideal simulation, where their energies agree with the exact ones to within $\sim\!10^{-4}\,t$, the whole hardware error is the error with which the device measures the optimal circuit. The same quantity in the aer\_noise calculation of Fig.~\ref{fig:ruido}, the measured energy compared with the ideal energy of the same parameters, is $2.6\times10^{-2}$ (armchair) and $2.2\times10^{-2}$ (zigzag), for circuits of the same size (two qubits, ansatz with one repetition, 1024 shots). The real devices therefore produce between $1.7$ and $2.9$ times the measurement error predicted by the noise model, unlike the single-qubit monolayer of Sec.~\ref{sec:res_monocapa}. The hardware deviations are also systematic rather than statistical: in seven of the eight bands the measured energy is shifted on average toward the center of the spectrum ($E=0$), the signature of incoherent errors that drive the prepared state toward the maximally mixed one, whereas in the noise simulation the mean shift of every band stays below $1.6\times10^{-2}\,t$. This is the expected effect of the error sources that the model leaves out and that act mainly on two-qubit gates: crosstalk, correlated errors and the drift of the calibration (Sec.~\ref{sec:metodos}).

\begin{figure*}[t]
  \centering
  \includegraphics[width=\textwidth]{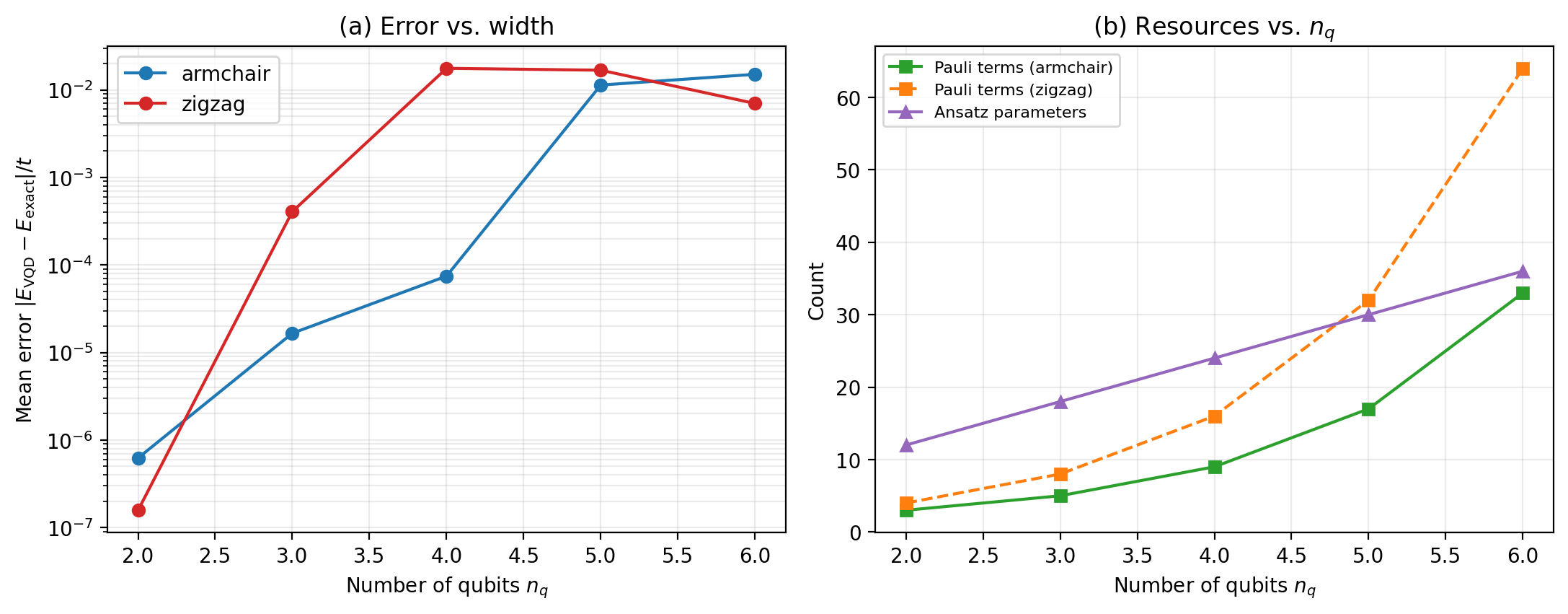}
  \caption{Scalability and error analysis (VQD of the four lowest bands; statevector mode throughout, SLSQP optimizer, EfficientSU2 ansatz with two repetitions). (a)~Mean error with respect to exact diagonalization versus the number of qubits $n_q$ (logarithmic scale). (b)~Resources versus $n_q$: number of Pauli terms of the Hamiltonian and number of ansatz parameters.}
  \label{fig:escalabilidad}
\end{figure*}

\begin{table*}[t]
  \centering
  \caption{Scalability with the number of qubits $n_q$ , for the VQD reconstruction of the four lowest bands on a grid of seven $k$ points in $[0,\pi]$, with an EfficientSU2 ansatz of two repetitions held fixed across all sizes. For each ribbon (subscript $a$: armchair; subscript $z$: zigzag) we give the number of ansatz parameters, the number of Pauli terms of the Hamiltonian, the number of cost-function evaluations and the mean error $\bar\epsilon/t$ with respect to exact diagonalization in the two execution modes. The statevector columns (SLSQP optimizer) measure exclusively the intrinsic limitation of the ansatz and of the classical optimization; the resource counts and the evaluation counts correspond to them. The aer\_noise columns repeat the same calculation on a device noise model. SPSA replaces SLSQP under noise. The hardware columns give the energy measured on real IBM Quantum processors (ibm\_fez, armchair; ibm\_kingston, zigzag) for $n_q=2$, the only width run on hardware. They correspond to Fig.~\ref{fig:hardware} and differ from the rest of the table in three respects: the hybrid strategy of Sec.~\ref{sec:metodos} (parameters optimized in ideal simulation, only the final measurement on the device), an ansatz with one repetition, and a grid of nine $k$ points in $[0,\pi]$.}
  \label{tab:escalabilidad}
  \begin{tabular}{ccccccccccccc}
    \toprule
    & & & \multicolumn{2}{c}{Pauli} & \multicolumn{2}{c}{Evals} &
    \multicolumn{2}{c}{$\bar\epsilon/t$ (statevector)} &
    \multicolumn{2}{c}{$\bar\epsilon/t$ (aer\_noise)} &
    \multicolumn{2}{c}{$\bar\epsilon/t$ (hardware)} \\
    \cmidrule(lr){4-5}\cmidrule(lr){6-7}\cmidrule(lr){8-9}\cmidrule(lr){10-11}\cmidrule(lr){12-13}
    $n_q$ & $N$ & Param. & $a$ & $z$ & $a$ & $z$ & $a$ & $z$ & $a$ & $z$ & $a$ & $z$ \\
    \midrule
    2 & 2  & 12 & 3  & 4  & 2\,566  & 3\,617   & $6.2\times10^{-7}$ & $1.6\times10^{-7}$ & $4.3\times10^{-2}$ & $3.1\times10^{-2}$ & $4.4\times10^{-2}$ & $6.5\times10^{-2}$ \\
    3 & 4  & 18 & 5  & 8  & 7\,315  & 11\,427  & $1.7\times10^{-5}$ & $4.1\times10^{-4}$ & $4.7\times10^{-2}$ & $8.3\times10^{-2}$ & -- & -- \\
    4 & 8  & 24 & 9  & 16 & 12\,207 & 30\,150  & $7.5\times10^{-5}$ & $1.8\times10^{-2}$ & $1.1\times10^{-1}$ & $1.0\times10^{-1}$ & -- & -- \\
    5 & 16 & 30 & 17 & 32 & 15\,823 & 60\,768  & $1.1\times10^{-2}$ & $1.7\times10^{-2}$ & $9.5\times10^{-2}$ & $5.7\times10^{-2}$ & -- & -- \\
    6 & 32 & 36 & 33 & 64 & 19\,170 & 121\,074 & $1.5\times10^{-2}$ & $7.0\times10^{-3}$ & $5.5\times10^{-2}$ & $6.1\times10^{-2}$ & -- & -- \\
    \bottomrule
  \end{tabular}
\end{table*}

Finally, we show a study of how the error and the resources scale as the ribbon width increases. We reconstruct by VQD the four lowest bands for $n_q=2,\dots,6$ (widths $N=2,\dots,32$), on a grid of seven $k$ points in $[0,\pi]$, with the SLSQP optimizer and an EfficientSU2 ansatz with two repetitions, held fixed across all sizes. The error $\bar\epsilon/t$ is the mean, over the four bands and the seven $k$ points, of the absolute deviation with respect to exact diagonalization.

The results show one favorable feature and one limiting feature. The favorable one is the scaling of the resources. The number of qubits grows logarithmically with the width, $n_q=\log_2(2N)$. The number of ansatz parameters is linear in $n_q$ ($6n_q$ for two repetitions); and thanks to the sparse character of the Bloch Hamiltonian, the number of Pauli terms grows approximately linearly with the width ($\sim N$ for armchair and $\sim 2N$ for zigzag) rather than with the worst-case exponential $4^{n_q}$ of a dense operator (Fig.~\ref{fig:escalabilidad}b). Since the measurement cost of a variational algorithm is governed by the number of Pauli terms to be estimated, the quantum resources per $k$ point remain modest as the physical system grows.

The limiting feature is the precision at fixed ansatz depth. As Fig.~\ref{fig:escalabilidad}a shows on a logarithmic scale, the mean error grows by four orders of magnitude across the range studied, from $\sim\!10^{-7}$--$10^{-6}$ at $n_q=2$ to $\sim\!10^{-2}$ for $n_q\gtrsim5$. The reason is that an ansatz of fixed expressivity has to cover a Hilbert space whose dimension doubles with every added qubit: with two repetitions the circuit has only $6n_q$ parameters, a number that grows linearly, and at some point the exact eigenvector ceases to be reachable.

Under noise the picture is different. The aer\_noise error lies between $3.1\times10^{-2}$ and $1.1\times10^{-1}$ at every size and shows no systematic trend with $n_q$. Up to $n_q=4$ it exceeds the statevector error by two to five orders of magnitude, so the intrinsic limitation of the ansatz is completely masked by the noise; only for $n_q\geq5$, where the ideal error has itself reached $\sim\!10^{-2}$, do the two come within an order of magnitude of each other. Separating the two contributions clarifies what happens. The optimization contribution (ideal energy of the parameters found under noise versus the exact one) varies between $2\times10^{-3}$ and $8\times10^{-2}$ without a clear trend. The measurement contribution (measured versus ideal energy of the same parameters) grows from $2.4$--$3.0\times10^{-2}$ at $n_q=2$--$3$ to $4.4$--$7.5\times10^{-2}$ for $n_q\geq4$, and it also changes in nature. For $n_q\leq3$ it is essentially statistical, with a mean shift of at most $1.2\times10^{-2}\,t$, whereas for $n_q\geq4$ the measured energies lie systematically above the ideal ones, by $4$--$7\times10^{-2}\,t$ on average, that is, shifted toward the center of the spectrum. This is the accumulation of gate errors in a circuit whose number of CX gates grows as $2(n_q-1)$: the larger the circuit, the closer the prepared state gets to the maximally mixed one.

The hardware columns, available for $n_q=2$, must be read with the caveats stated in the caption. The hardware error ($4.4\times10^{-2}$ armchair, $6.5\times10^{-2}$ zigzag) contains only the measurement contribution, and yet it is of the same order as the total aer\_noise error ($4.3\times10^{-2}$ and $3.1\times10^{-2}$), which contains both. Compared with the measurement contribution alone of the aer\_noise calculation ($3.0\times10^{-2}$ for both ribbons), the device is worse by a factor of $1.5$ (armchair) and $2.2$ (zigzag), in line with the analysis of Fig.~\ref{fig:hardware}.

Recovering the precision of the ideal simulation requires increasing the number of repetitions, at the cost of a deeper circuit. 

\begin{figure}[t]
  \centering
  \includegraphics[width=\columnwidth]{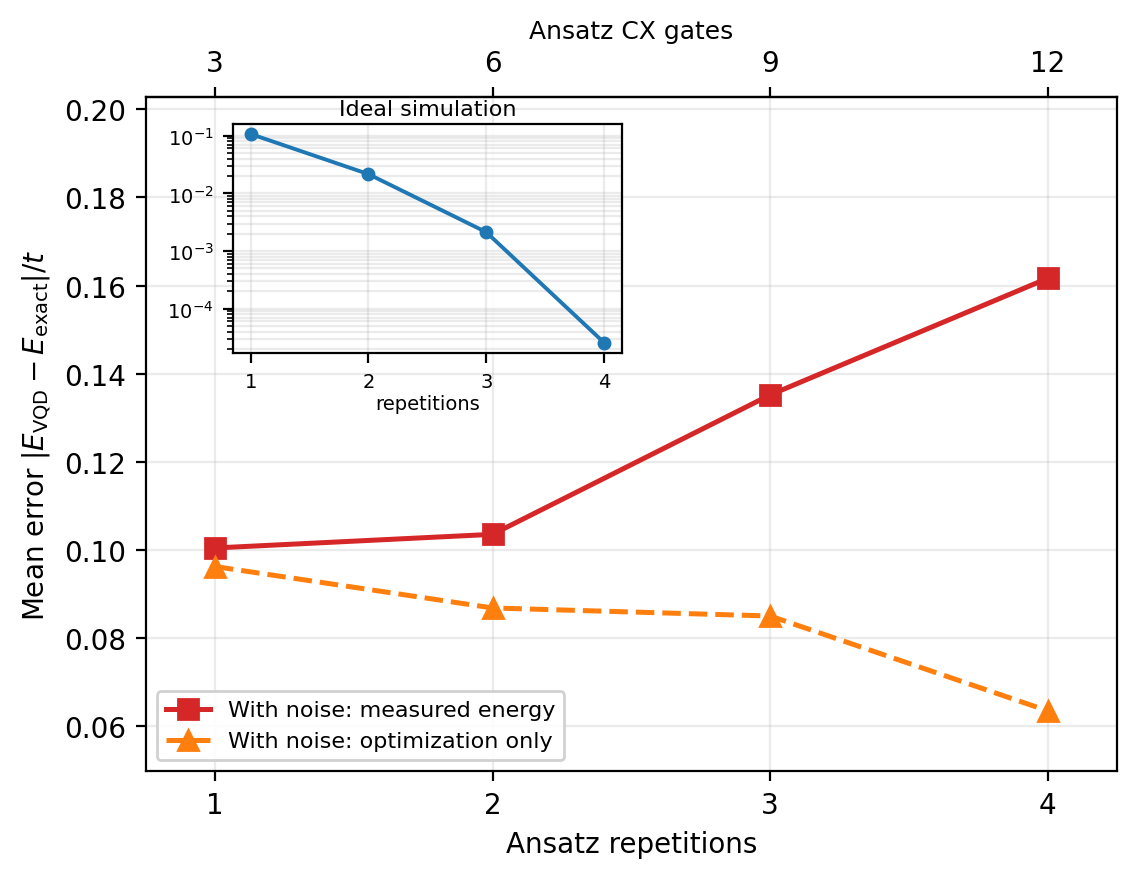}
  \caption{Effect of the ansatz depth on the accuracy of the reconstruction (zigzag ribbon, $n_q=4$, four lowest bands, seven $k$ points in $[0,\pi]$). Mean error with respect to exact diagonalization versus the number of repetitions of the EfficientSU2 ansatz, on a logarithmic scale; the upper axis gives the number of two-qubit gates of the corresponding circuit. Blue: ideal simulation (statevector, SLSQP). Red: energy measured under the device noise model (aer\_noise, SPSA, 1024 shots). Orange: ideal energy of the parameters obtained under noise, which separates the quality of the optimization from the error of the measurement.}
  \label{fig:reps}
\end{figure}

We repeated the reconstruction of the four lowest bands of the zigzag ribbon at $n_q=4$ for one to four repetitions of the ansatz, in statevector and aer\_noise modes, on the same grid of seven $k$ points and with the optimizer used throughout this work in each regime (SLSQP and SPSA, respectively); Fig.~\ref{fig:reps} collects the result. In statevector the gain is dramatic: the mean error falls monotonically from $1.1\times10^{-1}$ at one repetition to $2.6\times10^{-5}$ at four, three and a half orders of magnitude, as the circuit becomes expressive enough to reach the exact eigenvectors. Under noise the trend is the opposite. The measured error is essentially unchanged between one and two repetitions ($1.00\times10^{-1}$ and $1.04\times10^{-1}$) and then grows, to $1.35\times10^{-1}$ and $1.62\times10^{-1}$ at three and four, while the two-qubit gate count of the ansatz goes from three to twelve. Separating the two contributions shows that the ideal energy of the parameters found under noise does improve with depth ($9.6\times10^{-2}$ to $6.3\times10^{-2}$), so the optimization is not what fails; what degrades is the energy actually measured, because a deeper circuit accumulates more gate error at every evaluation. At four repetitions the two regimes differ by nearly four orders of magnitude, and the expressivity that ideal simulation rewards is precisely what the device punishes. The optimal depth on a noisy processor is therefore the shallowest one the problem tolerates, one or two repetitions, which is what we use throughout, whereas in ideal simulation deeper is always better. Since our noise model underestimates the noise of the real device (Sec.~\ref{sec:res_nanocintas}), the turnover on hardware can only come earlier. The two error mechanisms in the NISQ regime, the limited expressivity of a shallow ansatz and the flattening of the optimization landscape by noise, are well documented~\cite{McClean2018,WangNIBP2021}.

The two ribbons do not degrade alike. The zigzag ribbon loses precision earlier than the armchair one (it is already at $\bar\epsilon/t\approx1.8\times10^{-2}$ at $n_q=4$, where the armchair one is still at $7.5\times10^{-5}$) and it demands more work from the optimizer: at $n_q=6$ it requires $121\,074$ cost-function evaluations against $19\,170$ for the armchair ribbon, a factor of more than six. Both facts have the same origin: the lowest part of the zigzag spectrum is more crowded. This crowding is not due to the edge states, which lie at $E\approx0$, in the middle of the spectrum, and for $n_q\geq4$ are not among the four lowest bands. It comes from the bulk bands: at the zone boundary $k=\pi$ the zigzag Bloch Hamiltonian has the eigenvalue $-|t|$ with degeneracy $N-1$ (together with $+|t|$, also $(N-1)$-fold, and the two edge states at $E=0$), so the four lowest bands converge to the same energy as $k\to\pi$. On our grid, from $n_q=4$ on, their mean spacing is less than half that of the armchair ribbon ($0.15\,|t|$ against $0.37\,|t|$ at $n_q=4$, $0.011\,|t|$ against $0.038\,|t|$ at $n_q=6$). In such a spectrum the deflation penalty of Eq.~\eqref{eq:vqd} has to separate states whose energies differ by very little, the cost landscape becomes correspondingly flat, and the optimizer needs more evaluations to resolve one level from the next.

The same crowding explains why the zigzag error decreases between $n_q=4$ and $n_q=6$ (from $1.8\times10^{-2}$ to $7.0\times10^{-3}$), a decrease that does not mean that the reconstruction improves. When a variational state is not the exact eigenvector, it mixes the target level with its neighbors or settles on one of them, and the resulting energy error is bounded by the spacing between those levels. As the ribbon widens, the four lowest levels get closer together: in the zigzag ribbon their mean spacing falls from $0.15\,|t|$ at $n_q=4$ to $0.043\,|t|$ at $n_q=5$ and $0.011\,|t|$ at $n_q=6$. A worse wavefunction can therefore produce a smaller energy error. Measured in units of the level spacing, the error grows monotonically with $n_q$ in both ribbons (zigzag: $0.12$, $0.39$ and $0.62$ at $n_q=4$, $5$ and $6$; armchair: $0.08$ and $0.39$ at $n_q=5$ and $6$). The energy alone thus underestimates how much the reconstruction degrades as the spectrum becomes denser, which is precisely the limitation that motivates the wavefunction-level diagnostic of the next section.

\FloatBarrier

\section{Wavefunction quality}
\label{sec:containment}

The preceding sections show that the VQE/VQD workflow reproduces the energies of both systems with good precision. However, agreement in energy does not necessarily imply that the state prepared is the correct physical state. In this section we apply the subspace-containment diagnostic of Eqs.~\eqref{eq:containment}--\eqref{eq:containment2} to validate the reconstructed wavefunction. We proceed in the same order as before: first the monolayer, in two regimes statevector and aer\_noise; then the nanoribbons, in a three-qubit calculation whose parameters are optimized under aer\_noise; and we close with a joint discussion of the two systems.

\begin{figure*}[t]
  \centering
  \includegraphics[width=\textwidth]{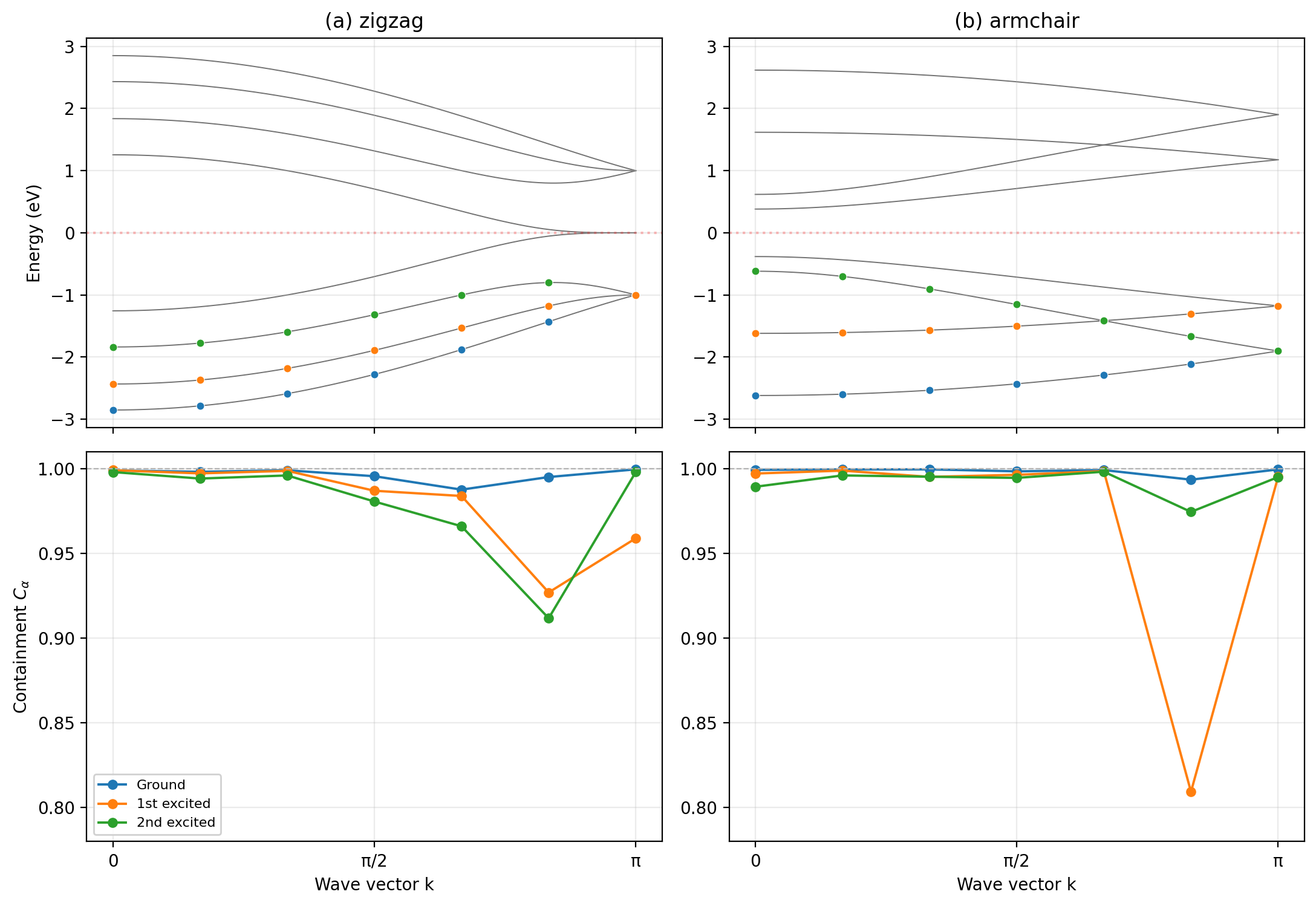}
  \caption{Subspace containment of the nanoribbons (three qubits, eight states), with the ansatz parameters optimized in the aer\_noise mode (generic $n_q$-qubit device, 1024 shots, SPSA, $n_r=3$ seeds per state) and the containment evaluated on the noiseless statevector of those parameters. Top row: exact band structure (gray lines) with the reconstructed levels highlighted; bottom row: containment $C_\alpha$ of each state (ground state and first two excited states) versus $k$, for the (a) zigzag and (b) armchair ribbons.}
  \label{fig:cont_nanocintas}
\end{figure*}

For monolayer graphene, evaluating the subspace containment of the valence and conduction bands along the path $\Gamma\!\to\!K\!\to\!M\!\to\!\Gamma$ demonstrates high wavefunction quality in both noiseless and simulated noise regimes. As a single-qubit system, its reconstruction is exceptionally robust, maintaining $C_\alpha \gtrsim 0.99$ for both bands across the Brillouin zone. At the degenerate Dirac point $K$, the manifold-projected containment approaches unity, confirming that the physical wavefunction is preserved in the two-dimensional system.

We turn now to the GNBs. In this case we have eight states instead of two, and three states to be reconstructed sequentially by VQD instead of one. Fig.~\ref{fig:cont_nanocintas} presents the containment of the ground state and the first two excited states of the three-qubit ribbons, with the parameters optimized in the aer\_noise mode. The reconstruction is good throughout, but not uniformly. The ground state is essentially exact in both ribbons ($C_\alpha \geq 0.988$ over the entire path). The excited states remain high, but fall off with the index of the state and with the position along the path: in the zigzag ribbon they stay at $C_\alpha \geq 0.91$, with the minimum ($0.927$ and $0.912$) at $k\approx 5\pi/6$, where the three lowest levels are already converging toward the threefold degenerate manifold at $E=-|t|$ that they form at $k=\pi$ and the spectrum becomes crowded; in the armchair ribbon they stay at $C_\alpha \geq 0.97$ except for the second excited state at $k\approx 5\pi/6$, which drops to $C_\alpha = 0.809$, leaking $17\%$ of its weight onto the eigenspace immediately above ($E=-1.031\,|t|$, a mere $0.27\,|t|$ away). This is the expected behavior of a deflation scheme: the last state of the sequence is the one that has accumulated the most penalty terms and inherits the imperfections of the two preceding ones, and it is therefore the first to leak toward its neighbor when the levels are close.

It should be stressed that these containment values are those of a converged optimization, in the sense of Sec.~\ref{sec:metodos}: with a single seed per state, the same calculation produced spurious collapses ($C_\alpha \approx 0.04$ at an isolated $k$ point of the armchair ribbon, $C_\alpha \approx 0.55$ at one point of the zigzag one) which did not reflect any physics but a state that had skipped a level, and which disappear on optimizing from three seeds. We record this because it is a practical hazard of VQD on noisy cost functions and, in the absence of a wavefunction-level diagnostic, it would be easy to mistake for a genuine effect.

The degenerate points deserve separate reading. At the zone boundary ($k=\pi$) of the zigzag ribbon, the three lowest states form a threefold degenerate manifold at $E=-|t|$. This is a bulk manifold, not the edge states: for $N=4$ the spectrum at $k=\pi$ is $\{-|t|^{(3)},\,0^{(2)},\,+|t|^{(3)}\}$, so the edge states at $E=0$ are the fourth and fifth levels, which are not reconstructed here. The three variational states jointly reside in that eigenspace, with individual containments of $1.000$, $0.998$ and $0.959$ over the manifold and a total close to unity. Almost all of the missing weight of the third state (3.8\%) goes to the next eigenspace, the edge-state manifold at $E=0$. The armchair ribbon has its own degeneracies, a twofold eigenspace at $k=2\pi/3$, where the two excited states give $0.999$ and $0.998$ over the manifold, and a further one at $k=\pi$. 

The contrast between the two systems is informative. In simulation, both are reconstructed correctly at the wavefunction level once the optimization has converged: the two-dimensional system stays at containment close to unity along the entire path, and the one-dimensional one keeps an essentially exact ground state and excited states above $0.9$, with a residual leakage that grows with the index of the state and concentrates where consecutive levels approach one another. The diagnostic therefore does not, by itself, condemn the variational reconstruction under a device noise model; what it does is sort the difficulty, ground state before excited states, isolated levels before crowded ones, and expose optimization failures that the energy would have disguised. 

\FloatBarrier

\section{Conclusions}
\label{sec:conclusions}

We have computed, within a single VQE/VQD variational workflow and in the $\pi$-electron approximation, the band structure of two graphene systems of different dimensionality: the two $\pi$ bands of the two-dimensional monolayer and the complete set of subbands of the one-dimensional zigzag- and armchair-edge nanoribbons, from ideal simulation to real quantum hardware. In ideal simulation, the method reproduces the bands of both systems with excellent precision against exact diagonalization. Under noise, the band structure is recovered with deviations of a few hundredths of $t$, with comparable contributions from finite sampling and from the noisy optimization. Execution on real IBM Quantum processors demonstrates the feasibility of the computation on current NISQ devices. With the parameters optimized in simulation, the error with which the device measures the optimal circuit agrees with the noise model for the single-qubit monolayer, but exceeds it by a factor of $1.5$--$3$ for the two-qubit circuits of the nanoribbons, where it also becomes systematic. The scalability study shows a favorable feature, the number of qubits grows logarithmically with the ribbon width, and a limiting one, the precision of a fixed-depth ansatz degrades as the dimension of the Hilbert space increases. Deepening the ansatz does not rescue that precision on a noisy device: a sweep over the number of repetitions improves the error by three and a half orders of magnitude in ideal simulation while making the energy measured under noise steadily worse, so that the useful depth is bounded from above by the hardware and not by the physics.

This work adds a validation of the wavefunction quality via subspace containment. Under a device noise model, the diagnostic shows that both systems are reconstructed correctly once the optimization has converged, containment close to unity for the monolayer, and above $0.9$ for the three states of the nanoribbons, and that the residual degradation is ordered: it grows with the index of the state in the deflation sequence and concentrates where consecutive levels approach one another. The diagnostic also proved able to expose a failure mode of VQD on noisy cost functions, a state settling on an eigenspace that is not its own, which the energy alone disguises and which is corrected by optimizing from several seeds. Furthermore, it naturally captures the degenerate manifolds of both systems.

The workflow validated here transfers to a fully on-device optimization without modification. In a broader perspective, the methodology validated here on non-interacting electron systems constitutes a controlled starting point toward the study of topological and strongly correlated phases in realistic two and one-dimensional materials.

\begin{acknowledgments}
J.B. acknowledges financial support from the Department of Education, Universities and Research of the Basque Government (Eusko Jaurlaritza) and the University of the Basque Country (UPV/EHU) under Grant No. 2122-26. U. A. acknowledges support from the Gipuzkoa Quantum
Programme financed by the Provincial Government of
Gipuzkoa with project number 2025-QUAN-000021-01
and 2025-QUAN-000037-01. U. A. acknowledges support from the ELKARTEK programme
financed by the Basque Goverment with project number
KK-2025/00079.
We acknowledge the use of IBM Quantum services for this work; the views expressed are those of the authors and do not reflect the official policy or position of IBM or the IBM~Quantum team.
\end{acknowledgments}

\FloatBarrier

\bibliographystyle{apsrev4-2}
\bibliography{refs}

@article{Wakabayashi2010,
  author  = {Wakabayashi, Katsunori and Sasaki, Ken-ichi and Nakanishi, Takeshi and Enoki, Toshiaki},
  title   = {Electronic states of graphene nanoribbons and analytical solutions},
  journal = {Science and Technology of Advanced Materials},
  volume  = {11},
  number  = {5},
  pages   = {054504},
  year    = {2010},
  doi     = {10.1088/1468-6996/11/5/054504}
}

@article{Son2006,
  author  = {Son, Young-Woo and Cohen, Marvin L. and Louie, Steven G.},
  title   = {Energy Gaps in Graphene Nanoribbons},
  journal = {Phys. Rev. Lett.},
  volume  = {97},
  pages   = {216803},
  year    = {2006},
  doi     = {10.1103/PhysRevLett.97.216803}
}

@article{Sherbert2022band,
  author  = {Kye Sherbert, Anooja Jayaraj and Marco Buongiorno Nardelli},
  title   = {Quantum algorithm for electronic band structures with local tight-binding orbitals},
  journal = {Sci. Rep.},
  number  = {12 (1):9867},
  year    = {2022},
  doi     = {10.1038/s41598-022-13627-x}
}

@misc{Exposito2026FQHE,
      title={Toward Quantum Utility in Correlated Topological Matter: Variational Preparation of Fractional Quantum Hall Manifolds},
      author={Sergio F. Exp{\'o}sito and Unai Aseginolaza and Ra{\'u}l Guerrero-Avil{\'e}s and Joaquim Jornet-Somoza and Francisco Guinea and Juan Borge},
      year={2026},
      eprint={2607.11380},
      archivePrefix={arXiv},
      primaryClass={quant-ph},
      url={https://arxiv.org/abs/2607.11380},
}

@article{Feynman1982,
  author  = {Feynman, Richard P.},
  title   = {Simulating physics with computers},
  journal = {Int. J. Theor. Phys.}, volume = {21}, pages = {467--488}, year = {1982},
  doi     = {10.1007/BF02650179}
}

@article{Peruzzo2014,
  author  = {Peruzzo, Alberto and McClean, Jarrod and Shadbolt, Peter and Yung, Man-Hong and Zhou, Xiao-Qi and Love, Peter J. and Aspuru-Guzik, Al{\'a}n and O'Brien, Jeremy L.},
  title   = {A variational eigenvalue solver on a photonic quantum processor},
  journal = {Nat. Commun.}, volume = {5}, pages = {4213}, year = {2014},
  doi     = {10.1038/ncomms5213}
}

@article{McClean2016,
  author  = {McClean, Jarrod R. and Romero, Jonathan and Babbush, Ryan and Aspuru-Guzik, Al{\'a}n},
  title   = {The theory of variational hybrid quantum-classical algorithms},
  journal = {New J. Phys.}, volume = {18}, pages = {023023}, year = {2016},
  doi     = {10.1088/1367-2630/18/2/023023}
}

@article{Kandala2017,
  author  = {Kandala, Abhinav and Mezzacapo, Antonio and Temme, Kristan and Takita, Maika and Brink, Markus and Chow, Jerry M. and Gambetta, Jay M.},
  title   = {Hardware-efficient variational quantum eigensolver for small molecules and quantum magnets},
  journal = {Nature}, volume = {549}, pages = {242--246}, year = {2017},
  doi     = {10.1038/nature23879}
}

@article{Preskill2018,
  author  = {Preskill, John},
  title   = {Quantum Computing in the {NISQ} era and beyond},
  journal = {Quantum}, volume = {2}, pages = {79}, year = {2018},
  doi     = {10.22331/q-2018-08-06-79}
}

@article{Bharti2022,
  author  = {Bharti, Kishor and Cervera-Lierta, Alba and Kyaw, Thi Ha and Haug, Tobias and Alperin-Lea, Sumner and Anand, Abhinav and Degroote, Matthias and Heimonen, Hermanni and Kottmann, Jakob S. and Menke, Tim and Mok, Wai-Keong and Sim, Sukin and Kwek, Leong-Chuan and Aspuru-Guzik, Al{\'a}n},
  title   = {Noisy intermediate-scale quantum algorithms},
  journal = {Rev. Mod. Phys.}, volume = {94}, pages = {015004}, year = {2022},
  doi     = {10.1103/RevModPhys.94.015004}
}

@article{McClean2018,
  author  = {McClean, Jarrod R. and Boixo, Sergio and Smelyanskiy, Vadim N. and Babbush, Ryan and Neven, Hartmut},
  title   = {Barren plateaus in quantum neural network training landscapes},
  journal = {Nat. Commun.}, volume = {9}, pages = {4812}, year = {2018},
  doi     = {10.1038/s41467-018-07090-4}
}

@article{Bauer2020,
  author  = {Bauer, Bela and Bravyi, Sergey and Motta, Mario and Chan, Garnet Kin-Lic},
  title   = {Quantum Algorithms for Quantum Chemistry and Quantum Materials Science},
  journal = {Chem. Rev.}, volume = {120}, pages = {12685--12717}, year = {2020},
  doi     = {10.1021/acs.chemrev.9b00829}
}

@article{Spall1992,
  author  = {Spall, J. C.},
  title   = {Multivariate stochastic approximation using a simultaneous perturbation gradient approximation},
  journal = {IEEE Trans. Autom. Control}, volume = {37}, pages = {332--341}, year = {1992},
  doi     = {10.1109/9.119632}
}

@article{BonetMonroig2023,
  author  = {Bonet-Monroig, X. and Wang, H. and Vermetten, D. and Senjean, B. and
             Moussa, C. and B{\"a}ck, T. and Dunjko, V. and O'Brien, T. E.},
  title   = {Performance comparison of optimization methods on variational quantum algorithms},
  journal = {Phys. Rev. A}, volume = {107}, pages = {032407}, year = {2023},
  doi     = {10.1103/PhysRevA.107.032407}
}

@article{WangNIBP2021,
  author  = {Wang, S. and Fontana, E. and Cerezo, M. and Sharma, K. and Sone, A. and
             Cincio, L. and Coles, P. J.},
  title   = {Noise-induced barren plateaus in variational quantum algorithms},
  journal = {Nat. Commun.}, volume = {12}, pages = {6961}, year = {2021},
  doi     = {10.1038/s41467-021-27045-6}
}

@article{morosan2012strongly,
  title={Strongly correlated materials},
  author={Morosan, Emilia and Natelson, Douglas and Nevidomskyy, Andriy H and Si, Qimiao},
  journal={Advanced Materials},
  volume={24},
  number={36},
  pages={4896--4923},
  year={2012},
  publisher={Wiley Online Library}
}

@article{mott1949basis,
  title={The basis of the electron theory of metals, with special reference to the transition metals},
  author={Mott, Nevill F},
  journal={Proceedings of the Physical Society. Section A},
  volume={62},
  number={7},
  pages={416--422},
  year={1949}
}

@article{hubbard1963electron,
  title={Electron correlations in narrow energy bands},
  author={Hubbard, John},
  journal={Proceedings of the Royal Society of London. Series A. Mathematical and Physical Sciences},
  volume={276},
  number={1365},
  pages={238--257},
  year={1963},
  publisher={The Royal Society London}
}

@article{imada1998metal,
  title={Metal-insulator transitions},
  author={Imada, Masatoshi and Fujimori, Atsushi and Tokura, Yoshinori},
  journal={Reviews of modern physics},
  volume={70},
  number={4},
  pages={1039},
  year={1998},
  publisher={APS}
}

@article{bednorz1986possible,
  title={Possible high T c superconductivity in the Ba- La- Cu- O system},
  author={Bednorz, J George and M{\"u}ller, K Alex},
  journal={Zeitschrift f{\"u}r physik B condensed matter},
  volume={64},
  number={2},
  pages={189--193},
  year={1986},
  publisher={Springer}
}

@article{scalapino2012common,
  title={A common thread: The pairing interaction for unconventional superconductors},
  author={Scalapino, Douglas J},
  journal={Reviews of Modern Physics},
  volume={84},
  number={4},
  pages={1383--1417},
  year={2012},
  publisher={APS}
}

@article{cao2018unconventional,
  title={Unconventional superconductivity in magic-angle graphene superlattices},
  author={Cao, Yuan and Fatemi, Valla and Fang, Shiang and Watanabe, Kenji and Taniguchi, Takashi and Kaxiras, Efthimios and Jarillo-Herrero, Pablo},
  journal={Nature},
  volume={556},
  number={7699},
  pages={43--50},
  year={2018},
  publisher={Nature Publishing Group UK London}
}

@article{pantaleon2023superconductivity,
  title={Superconductivity and correlated phases in non-twisted bilayer and trilayer graphene},
  author={Pantale{\'o}n, Pierre A and Jimeno-Pozo, Alejandro and Sainz-Cruz, H{\'e}ctor and Phong, V{\~o} Tiến and Cea, Tommaso and Guinea, Francisco},
  journal={Nature Reviews Physics},
  volume={5},
  number={5},
  pages={304--315},
  year={2023},
  publisher={Nature Publishing Group UK London}
}

@article{son2006energy,
  title={Energy gaps in graphene nanoribbons},
  author={Son, Young-Woo and Cohen, Marvin L and Louie, Steven G},
  journal={Physical review letters},
  volume={97},
  number={21},
  pages={216803},
  year={2006},
  publisher={APS}
}

@article{cao2017topological,
  title={Topological phases in graphene nanoribbons: junction states, spin centers, and quantum spin chains},
  author={Cao, Ting and Zhao, Fangzhou and Louie, Steven G},
  journal={Physical review letters},
  volume={119},
  number={7},
  pages={076401},
  year={2017},
  publisher={APS}
}

@article{groning2018engineering,
  title={Engineering of robust topological quantum phases in graphene nanoribbons},
  author={Gr{\"o}ning, Oliver and Wang, Shiyong and Yao, Xuelin and Pignedoli, Carlo A and Borin Barin, Gabriela and Daniels, Colin and Cupo, Andrew and Meunier, Vincent and Feng, Xinliang and Narita, Akimitsu and others},
  journal={Nature},
  volume={560},
  number={7717},
  pages={209--213},
  year={2018},
  publisher={Nature Publishing Group UK London}
}

@article{schollwock2011density,
  title={The density-matrix renormalization group in the age of matrix product states},
  author={Schollw{\"o}ck, Ulrich},
  journal={Annals of physics},
  volume={326},
  number={1},
  pages={96--192},
  year={2011},
  publisher={Elsevier}
}

@article{aryasetiawan1998gw,
  title={The GW method},
  author={Aryasetiawan, Ferdi and Gunnarsson, Olle},
  journal={Reports on progress in Physics},
  volume={61},
  number={3},
  pages={237--312},
  year={1998}
}

@article{georges1996dynamical,
  title={Dynamical mean-field theory of strongly correlated fermion systems and the limit of infinite dimensions},
  author={Georges, Antoine and Kotliar, Gabriel and Krauth, Werner and Rozenberg, Marcelo J},
  journal={Reviews of modern physics},
  volume={68},
  number={1},
  pages={13},
  year={1996},
  publisher={APS}
}

@article{bravyi2002fermionic,
  title={Fermionic quantum computation},
  author={Bravyi, Sergey B and Kitaev, Alexei Yu},
  journal={Annals of Physics},
  volume={298},
  number={1},
  pages={210--226},
  year={2002},
  publisher={Elsevier}
}

@article{jordan1928paulische,
  title={{\"U}ber das paulische {\"a}quivalenzverbot},
  author={Jordan, Pascual and Wigner, Eugene},
  journal={Zeitschrift f{\"u}r Physik},
  volume={47},
  number={9},
  pages={631--651},
  year={1928},
  publisher={Springer}
}

@book{nielsen2000quantum,
  title={Quantum computation and quantum information},
  author={Nielsen, Michael A and Chuang, Isaac L and others},
  volume={1},
  year={2000},
  publisher={Cambridge university press Cambridge}
}

@article{cerasoli2020quantum,
  title={Quantum computation of silicon electronic band structure},
  author={Cerasoli, Frank T and Sherbert, Kyle and S{\l}awi{\'n}ska, Jagoda and Buongiorno Nardelli, Marco},
  journal={Physical Chemistry Chemical Physics},
  volume={22},
  number={38},
  pages={21816--21822},
  year={2020},
  publisher={The Royal Society of Chemistry}
}

@article{cortes2022quantum,
  title={Quantum Krylov subspace algorithms for ground-and excited-state energy estimation},
  author={Cortes, Cristian L and Gray, Stephen K},
  journal={Physical Review A},
  volume={105},
  number={2},
  pages={022417},
  year={2022},
  publisher={APS}
}

@article{yu2025quantum,
  title={Quantum-centric algorithm for sample-based Krylov diagonalization},
  author={Yu, Jeffery and Moreno, Javier Robledo and Iosue, Joseph T and Bertels, Luke and Claudino, Daniel and Fuller, Bryce and Groszkowski, Peter and Humble, Travis S and Jurcevic, Petar and Kirby, William and others},
  journal={arXiv preprint arXiv:2501.09702},
  year={2025}
}

@article{zhang2024full,
  title={Full band-structure calculation of semiconducting materials on a noisy quantum processor},
  author={Zhang, Shaobo and Karim, Akib and Quiney, Harry M and Usman, Muhammad},
  journal={Physical Review A},
  volume={110},
  number={6},
  pages={062415},
  year={2024},
  publisher={APS}
}

@article{egger2021warm,
  title={Warm-starting quantum optimization},
  author={Egger, Daniel J and Mare{\v{c}}ek, Jakub and Woerner, Stefan},
  journal={Quantum},
  volume={5},
  pages={479},
  year={2021},
  publisher={Verein zur F{\"o}rderung des Open Access Publizierens in den Quantenwissenschaften}
}

@article{puig2025variational,
  title={Variational quantum simulation: A case study for understanding warm starts},
  author={Puig, Ricard and Drudis, Marc and Thanasilp, Supanut and Holmes, Zo{\"e}},
  journal={PRX Quantum},
  volume={6},
  number={1},
  pages={010317},
  year={2025},
  publisher={APS}
}

@article{ma2020quantum,
  title={Quantum simulations of materials on near-term quantum computers},
  author={Ma, He and Govoni, Marco and Galli, Giulia},
  journal={npj Computational Materials},
  volume={6},
  number={1},
  pages={85},
  year={2020},
  publisher={Nature Publishing Group UK, London}
}

@article{bassman2021simulating,
  title={Simulating quantum materials with digital quantum computers},
  author={Bassman Oftelie, Lindsay and Urbanek, Miroslav and Metcalf, Mekena and Carter, Jonathan and Kemper, Alexander F and de Jong, Wibe A},
  journal={Quantum Science \& Technology},
  volume={6},
  number={4},
  pages={043002},
  year={2021},
  publisher={IOP Publishing}
}

@article{cade2020strategies,
  title={Strategies for solving the Fermi-Hubbard model on near-term quantum computers},
  author={Cade, Chris and Mineh, Lana and Montanaro, Ashley and Stanisic, Stasja},
  journal={Physical Review B},
  volume={102},
  number={23},
  pages={235122},
  year={2020},
  publisher={APS}
}

@article{stanisic2022observing,
  title={Observing ground-state properties of the Fermi-Hubbard model using a scalable algorithm on a quantum computer},
  author={Stanisic, Stasja and Bosse, Jan Lukas and Gambetta, Filippo Maria and Santos, Raul A and Mruczkiewicz, Wojciech and O’Brien, Thomas E and Ostby, Eric and Montanaro, Ashley},
  journal={Nature communications},
  volume={13},
  number={1},
  pages={5743},
  year={2022},
  publisher={Nature Publishing Group UK, London}
}

@article{kim2023evidence,
  title={Evidence for the utility of quantum computing before fault tolerance},
  author={Kim, Youngseok and Eddins, Andrew and Anand, Sajant and Wei, Ken Xuan and Van Den Berg, Ewout and Rosenblatt, Sami and Nayfeh, Hasan and Wu, Yantao and Zaletel, Michael and Temme, Kristan and others},
  journal={Nature},
  volume={618},
  number={7965},
  pages={500--505},
  year={2023},
  publisher={Nature Publishing Group UK, London}
}

\end{document}